\documentclass[aps,amsmath,amssymb,twocolumn,floatfix,pra,footinbib,superscriptaddress,longbibliography]{revtex4-2}
\usepackage{graphicx} % Required for inserting images

\usepackage{amsmath}
\usepackage{physics}
\usepackage{mleftright}
\usepackage{booktabs}
\usepackage{siunitx}

\usepackage[colorlinks]{hyperref}
\hypersetup{%
	plainpages=true,
	breaklinks=true,%not default in dvips mode, so we must specify
	hypertexnames=false,%not ideal, but needed when pagenums duplicate (`i' vs. `1')
	pageanchor=true,
	colorlinks=true,
	linkcolor={[RGB]{32,119,180}},
	citecolor={[RGB]{255,127,15}},
	urlcolor={[RGB]{32,119,180}},
	anchorcolor={black}
}

\let\textdotlessi\i                 % keep the text dotless i (needed by \'i, í in references)
\renewcommand*{\i}{\ifmmode\mathrm{i}\else\textdotlessi\fi}      % Imaginary unit (math); dotless i (text)
\newcommand*{\e}{\mathrm{e}}        % Euler's number
\newcommand*{\rd}{\mathrm{d}}        % Differential

\newcommand{\figref}[1]{\mbox{Fig.~\ref{#1}}}
\newcommand{\tabref}[1]{\mbox{Table~\ref{#1}}}
\newcommand{\secref}[1]{\mbox{Section~\ref{#1}}}
\newcommand{\sisecref}[1]{\mbox{Supplementary Section~\ref{#1}}}

\renewcommand{\eqref}[1]{\mbox{Eq.~(\ref{#1})}}

\newcommand{\figpanel}[2]{Fig.~\hyperref[#1]{\ref*{#1}(#2)}}
\newcommand{\figpanels}[3]{Fig.~\hyperref[#1]{\ref*{#1}(#2)--(#3)}}
\newcommand{\figpanelNoPrefix}[2]{\hyperref[#1]{\ref*{#1}(#2)}}

\begin{document}

\title{Modeling couplings and quench dynamics of hundred-qubit quantum processors}

\author{Simon Pettersson Fors}
\email{simon.pettersson.fors@chalmers.se}
\affiliation{Department of Microtechnology and Nanoscience, Chalmers University of Technology, 412 96 Gothenburg, Sweden}

\author{Eric Djäknegren}
\affiliation{Department of Physics and Astronomy, Chalmers University of Technology, 412 96 Gothenburg, Sweden}

\author{Jorge Fern\'andez-Pend\'as}
\affiliation{Department of Microtechnology and Nanoscience, Chalmers University of Technology, 412 96 Gothenburg, Sweden}

\author{Anton Frisk Kockum}
\affiliation{Department of Microtechnology and Nanoscience, Chalmers University of Technology, 412 96 Gothenburg, Sweden}

\date{\today}

\begin{abstract}
    Quantum processors now operate beyond the reach of exact classical simulation, yet their design and verification rest on precise models of the device. Understanding which properties of quantum processors can still be predicted on classical computers is central both to building these machines and to delimiting claims of quantum advantage. Here we address this question at the device level, using continuous unitary transformations to derive effective Hamiltonians non-perturbatively for two-dimensional superconducting processors at a cost polynomial in qubit number. For a 96-transmon array with tunable couplers, we extract the effective qubit couplings, state-dependent interactions (e.g., ZZ couplings), and eigenstate locality that guide processor design. This computation parallelizes over tensor contractions and thus takes an hour at most per design point on a single GPU, enabling rapid design iterations of the full array. Combining these effective models with semiclassical sampling, we then predict the quench dynamics of 100 qubits and 180 couplers at half filling, a sector of dimension $10^{29}$, and validate the approach against exact dynamics of smaller arrays. We show that our approach captures relaxation, thermalization, and transport, but not the bitstring statistics underlying cross-entropy benchmarks. The predicted transport carries a fingerprint of the device: under a staggered detuning, excitations cross over from ballistic to diffusive spreading, slow down at a doublon resonance, and diffuse at different rates on each sublattice. Our results chart which observables lie within classical reach and offer a practical tool for designing future devices.
\end{abstract}

\maketitle

\section{Introduction}

Quantum processors comprising hundreds of physical qubits now stand at the threshold of exploring quantum many-body dynamics beyond the reach of classical computation~\cite{bohnetQuantumSpinDynamics2016, GoogleQuantumAI2025, andersenThermalizationCriticalityAnalogue2025a, manovitzQuantumCoarseningCollective2025, Haghshenas2026}. However, translating raw qubit counts into useful computational scale is severely hampered by microscopic hardware non-idealities, such as parasitic couplings~\cite{sungRealizationHighFidelityCZ2021}, control-signal crosstalk, and unwanted drive dynamics, that degrade gate fidelities and distort target Hamiltonians~\cite{mohseniHowBuildQuantum2024, klimovOptimizingQuantumGates2024, GoogleQuantumAI2025QEC}. Unraveling and mitigating these physical constraints demands precise microscopic modeling. Yet, this creates a fundamental dilemma: classically predicting the dynamics of a noisy quantum processor is itself a quantum many-body problem of exponential complexity. As brute-force exact simulations rapidly fail at scale, there is a critical need for scalable, non-exact modeling frameworks capable of capturing multi-qubit coherent imperfections to guide the design, optimization, and verification of large-scale quantum hardware. The aim of such frameworks is not to compete with intermediate processors on classically hard computations, but to reveal how the device-level physics changes as processors are scaled up.

To address this challenge, physical hardware modeling has adapted numerical tools from quantum many-body physics, each relying on distinct physical approximations. Perturbative frame transformations, such as Schrieffer–Wolff expansions~\cite{schriefferRelationAndersonKondo1966,bravyiSchriefferWolffTransformationQuantum2011}, reliably yield effective Hamiltonians for weakly coupled static subsystems, but break down in strongly interacting regimes and for strong, near-resonant, or aperiodic control drives. Tensor network methods bypass perturbative assumptions; in particular, one-dimensional matrix product states have shown promise for modeling static couplings~\cite{gonzalez-garciaMultitargetDMRGXAlgorithm2026,schollwockDensitymatrixRenormalizationGroup2011}. However, tensor networks remain strictly bound by the growth of quantum entanglement. As correlations propagate in accordance with Lieb–Robinson bounds~\cite{liebFiniteGroupVelocity1972,bravyiLiebRobinsonBoundsGeneration2006}, required bond dimensions scale exponentially~\cite{schuchEntropyScalingSimulability2008}, rendering two-dimensional architectures and long-time dynamics intractable, a bottleneck further compounded by severe parallelization limits in tensor network contractions. Alternative approaches fare no better: linked-cluster expansions~\cite{rigolNumericalLinkedClusterApproach2006,mallayyaQuantumQuenchesRelaxation2018} are limited to short times and small clusters, quantum Monte Carlo~\cite{troyerComputationalComplexityFundamental2005, cohenTamingDynamicalSign2015} suffers from a dynamical sign problem that makes the cost grow exponentially with evolution time, and non-equilibrium dynamical mean-field theory~\cite{georgesDynamicalMeanfieldTheory1996, aokiNonequilibriumDynamicalMeanfield2014} is limited by the cost and stability of its real-time impurity solvers. Ultimately, these widespread failure modes are not mere technical hurdles, but direct manifestations of the fundamental computational hardness of device-level dynamics, highlighting the pressing need for entirely different modeling paradigms tailored to large-scale quantum hardware.

Overcoming these limitations requires a method tailored to the specific modeling challenges of quantum hardware, demanding capture of non-perturbative effective couplings, rapid simulation turnaround in design iterations, and native capacity for scaled-up two-dimensional systems. In this article, we introduce a method that meets these requirements by systematically deriving effective Hamiltonians that capture the essential physics of large systems. The method uses continuous unitary transformations (CUTs)~\cite{chuProjectedGradientMethod1990, brockettDynamicalSystemsThat1991, glazekRenormalizationHamiltonians1993, wegnerFlowequationsHamiltonians1994a, kehreinFlowEquationApproach2006}, also referred to as ``flow equations'', to construct renormalization flows towards the effective models, which scale polynomially in system size as a result of finite operator expansions. 

How to derive the flow equations in terms of the coefficients of the operator expansions was recently established for fermions in Refs.~\cite{thomsonLocalIntegralsMotion2023, thomsonUnravellingQuantumDynamics2024}, and we here make the extension to bosonic models. Bosons introduce additional multi-excitation resonances, in turn worsening the convergence properties of standard CUT implementations in ordered and non-perturbative regimes. We solve this problem by creating a CUT with a non-vanishing generator in ordered regimes, resulting in a substantially faster and more stable method. Leveraging these improvements, we apply CUTs to arrays of superconducting qubits with tunable couplers. For a 96-transmon array, we derive block-diagonal and diagonal effective Hamiltonians, from which we extract the effective qubit--qubit couplings, state-correlated (ZZ) couplings, and spatial eigenstate locality that guide processor design. On a single GPU, this computation takes an hour at most per design point because of the polynomial cost with system size and parallelizes over tensor contractions.

We then ask to which extent such effective models can predict other observables in a superconducting quantum processor, by studying the example of quench dynamics in analog simulators. Approximate diagonalization, the most direct route to the dynamics, fails for transmon arrays. The transmons act as hard-core bosons which require non-linear terms in the flowed operators that are as large as the linear ones at weak disorder. The truncated flow becomes accurate only for detunings beyond the anharmonicity, whereas analog simulators are operated well below it. We therefore use CUTs only for block diagonalization and evolve the resulting effective qubit model with a semiclassical Monte Carlo method, combining an excitation-cluster expansion~\cite{rigolNumericalLinkedClusterApproach2006, yangQuantumManybodyTheory2008} at early times with the discrete truncated Wigner approximation (DTWA)~\cite{polkovnikovPhaseSpaceRepresentation2010, schachenmayerManyBodyQuantumSpin2015} at intermediate and late times. For a device of 280 transmons with 50 excitations, this approach predicts the ballistic dynamics at resonance to sublattice-resolved diffusion under a staggered detuning, a transport crossover that offers an experimental test of the device model. It yields local densities and low-order correlators, but falls short at predicting bitstring probabilities. Our results therefore do not challenge cross-entropy-based claims of quantum advantage, but reveal observables which remain predictable on classical computers. As quantum processors grow beyond exact simulation, we expect effective models like these to guide their design and verification, by showing which properties can be predicted classically and which only the hardware itself can reveal.

\section{Continuous unitary transformations}
\begin{figure}
    \centering
    \includegraphics[width=\linewidth]{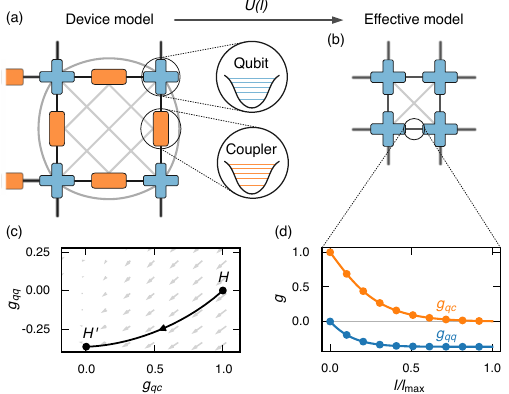}
    \caption{From device to effective models with continuous unitary transformations. 
    (a) The quantum hardware consists of qubits (blue crosses) coupled capacitively (black lines) to couplers (orange rectangles), and stray non-nearest-neighbor couplings (gray lines). The relative strength of the couplings is represented by the blackness of the lines. 
    (b) Effective qubit model after elimination of the coupler degrees of freedom through the continuous unitary transformation $U(l)$. 
    (c) The vector field given by the flow equation. The solution of the flow equation is a trajectory that connects the device model $H$ and the effective model $H'$. 
    (d) The flow of qubit-coupler and qubit-qubit couplings $g_\mathrm{qc}$ and $g_\mathrm{qq}$ during the integration of the flow equation to flow time $l_\mathrm{max}$. While $g_\mathrm{qc}$ is eliminated, an effective $g_\mathrm{qq}$ builds up. The gray line marks $g = 0$.}
    \label{fig:1}
\end{figure}

We consider quantum hardware with device (or microscopic) Hamiltonians that can be expanded in an operator basis of bosonic ladder operators:
\begin{equation}
\begin{aligned}
    H = &\sum_{ij} H_{ij} a_i^\dag a_j + \sum_{ijkl} H_{ijkl} a_i^\dag a_j^\dag a_k a_l + \\
    &\sum_{ij} V^\mathrm{nec}_{ij} \mleft( a_i^\dag a_j^\dag + \mathrm{H.c.} \mright) + \sum_{ijlk} V^\mathrm{nec}_{ijkl} \mleft( a_i^\dag a_j^\dag a_k^\dag a_l + \mathrm{H.c.} \mright).
    \label{eq:device_hamiltonian}
\end{aligned}
\end{equation}
This Hamiltonian is universal for many different bosonic systems, including, e.g., the Bose--Hubbard model~\cite{fisherBosonLocalizationSuperfluidinsulator1989}, neutral atoms in optical lattices~\cite{jakschColdBosonicAtoms1998, blochManybodyPhysicsUltracold2008}, and superconducting-qubit arrays~\cite{kochChargeinsensitiveQubitDesign2007, krantzQuantumEngineerGuide2019}, as in \figpanel{fig:1}{a}. The first two terms in \eqref{eq:device_hamiltonian} contain the excitation-conserving (ec) operators, while the last two terms are non-excitation conserving (nec), which are needed to properly model interacting superconducting qubits. We assume a finite array with $N$ bosonic modes, but impose no other constraints on the dimensionality nor the connectivity of the modes, captured by the coefficient tensors $H_{ij}$, $H_{ijkl}$, etc. Instead, the complexity of the model is determined by its number of parameters, in turn set by two axes: $N$ and the length $2K$ of strings of ladder operators corresponding to $K$-body interactions. Limiting the length of these strings corresponds to a truncation of the operator basis. For example, the one- and two-body couplings, describing the excitation-conserving interaction strengths between one or two excitations, are stored in the tensors $H_{ij}$ and $H_{ijkl}$, requiring $\mathcal{O}(N^2)$ and $\mathcal{O}(N^4)$ memory space, respectively. $K$-order interactions described by $K$-body couplings are stored in tensors of size $\mathcal{O}(N^{2K})$, thus growing polynomially with $N$ and exponentially with $K$, together setting the size of the operator basis.

The device Hamiltonian in \eqref{eq:device_hamiltonian} is not a unique description of the system. As a consequence of unitarity, there is an infinite family of Hamiltonians $H' = U H U^\dag$ related through unitary frame transformations $U$ that all result in the same physics. It generally has an exponential cost in the number of modes to compute $H'$ to arbitrary precision. For this reason, we search for effective models that do not capture the full complexity of the exact physics but achieve a sufficiently accurate approximation at a lower computational cost. A key idea in these approximations is to exploit scale hierarchies to eliminate weakly coupled degrees of freedom. Examples of such effective models are $U(1)$-symmetric, i.e., excitation-conserving, models where the non-excitation-conserving couplings in \eqref{eq:device_hamiltonian} have been eliminated, and effective qubit models through elimination of the couplers. Another example is approximately diagonal models where the time evolution of the system is straightforwardly obtained from exponentiation of the diagonal Hamiltonian.

Continuous unitary transformations provide a framework for systematically mapping a Hamiltonian onto an effective model~\cite{chuProjectedGradientMethod1990,brockettDynamicalSystemsThat1991, glazekRenormalizationHamiltonians1993,wegnerFlowequationsHamiltonians1994a,kehreinFlowEquationApproach2006}. In this scheme, the total unitary transformation is built from a continuous product of infinitesimal transformations $U(l) = \prod_{l'=0}^l \exp[\eta(l')\rd l']$, governed by an anti-Hermitian flow generator $\eta(l')$ tailored to the target physics. In the continuum limit $\rd l' \to 0$, $U(l)$ is a time-ordered exponential in the fictitious ``flow time'' $l$, giving a continuum of unitarily equivalent Hamiltonians $H(l) = U(l) H(0) U^\dag (l)$. The continuum stretches from the initial device Hamiltonian at $l = 0$ to the effective Hamiltonian $H'$ at the end of the flow, canonically at $l \to \infty $. Differentiating $H(l)$ with respect to $l$ gives a practical way to solve for the transformation, resulting in the celebrated ``flow equation''~\cite{wegnerFlowequationsHamiltonians1994a}
\begin{equation}
    \frac{\rd H}{\rd l} = \mleft[\eta(l), H(l) \mright].
\end{equation}

The right-hand side of the flow equation defines a vector field [see \figpanel{fig:1}{c}] which governs the flow of the Hamiltonian coefficients. Solutions to the flow equation correspond to trajectories in the vector field with the effective Hamiltonian as a fixed point, i.e., a steady-state solution $\rd H' / \rd l = 0$. The vector field, and especially the stability of its fixed points, depend on the choice of the generator. Similar to the Hamiltonian, we expand the generator in a ladder-operator basis: 
\begin{equation}
\begin{aligned}
    \eta = &\sum_{ij} \eta_{ij} a_i^\dag a_j + \sum_{ijkl} \eta_{ijkl} a_i^\dag a_j^\dag a_k a_l + \\
    &\sum_{ij} \eta^\mathrm{nec}_{ij} \mleft( a_i^\dag a_j^\dag - \mathrm{H.c.} \mright) + \sum_{ijlk} \eta^\mathrm{nec}_{ijkl} \mleft( a_i^\dag a_j^\dag a_k^\dag a_l - \mathrm{H.c.} \mright).
    \label{eq:generator_definition}
\end{aligned}
\end{equation}
The generator is further set by specifying a protocol for computing the coefficient tensors $\eta_{ij}(l)$, $\eta_{ijkl}(l)$, etc. Crucially, since we know the commutation relations for the ladder operators, the commutator in the flow equation is efficiently computed as tensor contractions between the different coefficient tensors at a cost polynomial in system size $N$.

In this work, we have developed a stable and scalable generator that achieves significantly faster convergence than tested alternatives for near-degenerate and ordered bosonic systems. The problem with other canonical generators, like the Wegner generator $\eta = [H_0, V] \propto E_i - E_j$, is that it is suppressed near degenerate energies $E_i - E_j \approx 0$, in turn causing vanishing-gradient problems and unstable convergence for integration of the flow equation. Here, $H_0(l) = \sum_{i=j} H_{ij}(l) a_i^\dag a_j + \sum_{(ij)=(kl)} H_{ijkl}(l) a_i^\dag a_j^\dag a_k a_l$ is the diagonal part of the Hamiltonian, where $(ij)=(kl)$ denotes equal index pairs independent of ordering. $E_i$ and $E_{ij}$ respectively denote the eigenenergies $H_0(l) \ket{1_i} = E_i(l) \ket{1_i}$ and $H_0(l) \ket{1_i,1_j} = E_{ij}(l) \ket{1_i, 1_j}$ of the eigenstates $\ket{1_i}$ and $\ket{1_i, 1_j}$, representing one (two) excitation(s) in modes $i$ and $j$ (if $i = j$). The remaining off-diagonal couplings are captured by $V(l) = H(l) - H_0(l)$. Given this problem, recent works~\cite{thomsonUnravellingQuantumDynamics2024} have successfully applied so-called ``scrambling transformations'' to perturb the flow trajectories away from unstable fixed points in fermionic models. However, bosonic systems involve two-excitation degeneracies forbidden for fermions by the Pauli exclusion principle, leading to additional unstable fixed points. These degeneracies are not avoided through scrambling, resulting in convergence issues. 

Rather than dealing with the shortcomings of the Wegner generator, we alleviate the convergence difficulties through extending the so-called sign generator~\cite{mielkeFlowEquationsBandmatrices1998,knetterPerturbationTheoryFlow2000} to operator expansions, having the advantage of being non-vanishing near degeneracies. The sign generator for excitation-conserving couplings is set by the coefficient tensors
\begin{subequations}
\begin{align}
    \eta_{ij}(l) &= \mathrm{asgn}(E_i-E_j) V_{ij}(l), \\
    \eta_{ijkl}(l) &= \mathrm{asgn}(E_{ij} - E_{kl}) V_{ijkl}(l),
\end{align}
\label{eq:eta_coefficients}
\end{subequations}
The non-vanishing property is given by the anti-symmetric sign function defined as
\begin{equation}
    \mathrm{asgn}(x_{IJ}) = 
\begin{cases}
    +1 &\text{for } x_{IJ} \geq 0, \\
    -1 &\text{for } x_{IJ} < 0, \\
    - \mathrm{sgn}(x_{JI}) &\text{for } I \leftrightarrow J, 
\end{cases}
\end{equation}
where $I$ and $J$ are an index or an index pair. Hence, even at degenerate energies, the sign generator creates a flow away from degeneracies and towards zero off-diagonal couplings, as can be seen from the flow equation, e.g., for the one-body couplings: $\rd H_{ii} / \rd l = \sum_j \mathrm{asgn}(E_i-E_j) V_{ij} V_{ji}$ and $\rd V_{ij} / \rd l = - \abs{E_i - E_j} V_{ij}$, respectively. The sign generator is straightforward to generalize beyond two-body couplings and to non-excitation-conserving couplings. Compared to the scrambling transformation, the sign generator has the advantage of not requiring any external monitoring loops, and we find it to converge ten times faster near degeneracies without any performance loss for non-degenerate systems (Methods; \figref{fig:validation}).

\begin{figure*}
    \centering
    \includegraphics[width=\linewidth]{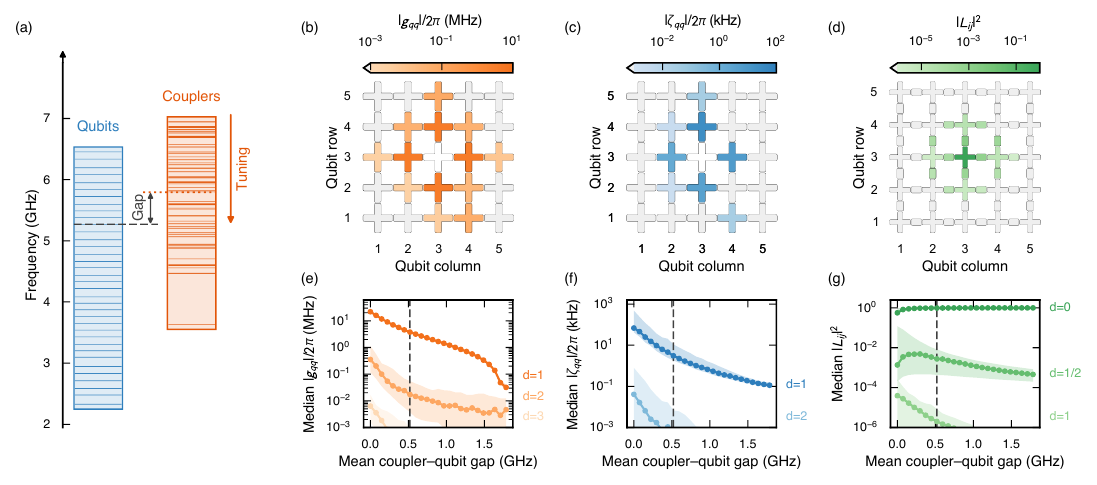}
    \caption{Design metrics for a 96-transmon array of $6 \times 6$ qubits and 60 couplers, as the couplers are tuned towards the qubits (Methods; \secref{sec:design_sweep}). The weakest coupled row 6 and column 6 are omitted in (b)--(d) to highlight key details. 
    (a) Sketch of the sweep. The 36 qubit frequencies (blue lines) form a disordered band, here spanning 2.3--\SI{6.5}{\giga\hertz} with a minimum separation of \SI{0.1}{\giga\hertz}, and the 60 coupler frequencies (orange lines) start well above it, at the frequency where the second-order qubit--qubit coupling vanishes, and are tuned down towards the higher-frequency qubit of each coupled qubit pair. The gap is the difference between the mean coupler frequency and the mean frequency of these target qubits (dashed). 
    (b)--(d) The effective qubit--qubit couplings $|g_\mathrm{qq}|/2\pi$, ZZ couplings $|\zeta_\mathrm{qq}|/2\pi$, and squared locality overlaps $|L_{ij}|^2$ between the central qubit in row 3 and column 3 [white in (b) and (c)] and its $5 \times 5$ subset qubit and coupler neighbors in rows and columns 1--5 at a mean gap of \SI{0.52}{\giga\hertz}. In (d), qubits are located at integer sites and couplers between the qubits they connect. Values below the colorbar range are shown in light gray. 
    (e)--(g) Median of the same quantities over all pairs of sites at distance $d$ as a function of the mean gap, with the shaded area spanning the 25th to 75th percentile. The distance is the Manhattan distance on the qubit lattice, in units where nearest-neighbor qubits have $d = 1$ and a qubit and its nearest coupler have $d = 1/2$; $d = 0$ is the overlap of a state with itself. The dashed vertical lines mark the gap used in (b)--(d), and distances not shown lie below the plotted range. Results are for a single realization of the device parameters (qubit--coupler coupling \SI{50}{\mega\hertz}, direct qubit--qubit coupling \SI{1.3}{\mega\hertz}, all with \SI{5}{\percent} disorder when sampled) and took less than 20~GPU-hours to compute.}
    \label{fig:design}
\end{figure*}

\section{Design metrics in superconducting processors}
\label{sec:design}

Like other renormalization group methods, the flow equation continually introduces higher-order interactions into the effective Hamiltonian, driven by non-linear generator terms containing three or more ladder operators. Maintaining computational tractability requires truncating this expansion at a fixed operator rank, effectively imposing a self-similarity constraint on the Hamiltonian. Fortunately, this truncation is often benign for different reasons. In the presence of scale separation $E_{ij} \gg |V^\mathrm{nec}_{ij}|$, as in transmon systems, the flow to higher-order non-excitation-conserving interactions, such as those in \eqref{eq:device_hamiltonian}, is heavily suppressed (Methods; \figref{fig:nec_validation}). For excitation-conserving terms, similar scale separations break down in non-perturbative parameter regimes. Yet, the excitation-conserving flow has the peculiar property that it is strictly unidirectional from few- to many-body couplings (Supplementary Information; \secref{SI:unidirectional}). Consequently, higher-order couplings exert no feedback on lower-order terms, allowing few-excitation properties to be calculated exactly or with high precision despite discarding the higher-order flow.

Exploiting this precise agreement in the low-excitation physics enables a systematic, multi-stage strategy for extracting critical hardware metrics in superconducting-qubit arrays --- namely, effective qubit couplings, state-dependent energy shifts like ZZ couplings, and spatial eigenstate locality. These three design metrics are key to understanding gate and noise performance~\cite{yanTunableCouplingScheme2018, sungRealizationHighFidelityCZ2021, pettersonforsComprehensiveExplanationZZ2024, krinnerBenchmarkingCoherentErrors2020, gonzalez-garciaMultitargetDMRGXAlgorithm2026}, and are illustrated in \figref{fig:design} for a $6 \times 6$ qubit array with couplers, i.e., a 96-transmon array consisting of 36 qubit and 60 coupler transmons. Because the flow to non-excitation-conserving interactions in \eqref{eq:device_hamiltonian} is suppressed, we eliminate it first with negligible truncation error (Methods; \figref{fig:nec_validation}). This initial transformation maps the raw device Hamiltonian into an effective $U(1)$-symmetric frame $H^\mathrm{ec}$ where the total excitation number is strictly conserved. From this unified excitation-conserving baseline, the subsequent flow is tailored depending on whether one seeks an effective interacting qubit model or a fully diagonalized spectrum.

To extract effective qubit exchange couplings while retaining coherent qubit dynamics, integral for simulating gate control and microwave crosstalk, we set $V_{ij} = 0$ and $V_{ijkl} = 0$ between qubit modes in \eqref{eq:eta_coefficients} for the generator to eliminate the exchange couplings involving coupler modes, while preserving qubit-qubit exchange. Eliminating the couplers yields an effective $U(1)$-symmetric qubit Hamiltonian:
\begin{equation}
    H^q = \sum_{i,j\in Q} H^q_{ij} q_i^\dag q_j + \sum_{ijkl \in Q} H^q_{ijkl} q_i^\dag q_j^\dag q_k q_l + \mathcal{O}(q^4, q^{\dag 3} q^3),
\label{eq:qubit_hamiltonian}
\end{equation}
where $Q$ indexes the active qubit array with effective creation operators $q_i^\dag$ and we have assumed that the couplers occupy their ground states. This assumption is not necessary; the Hamiltonian coefficients are coupler-state dependent without it. From \eqref{eq:qubit_hamiltonian}, the effective single-excitation exchange coupling $g_{ij}$ between qubits $i$ and $j$ ($i \neq j$) is identified directly as
\begin{equation}
    g_{ij} = H^q_{ij} \quad (i \neq j).
\end{equation}
While $g_{ij}$ governs primary excitation hopping, higher-order four-body terms such as $H^q_{ijkl}$ naturally capture state-dependent spectator shifts; a key error source for two-qubit-gate pulses in multi-qubit architectures. For the 96-transmon array in \figref{fig:design}, the effective coupling between qubits is largest for nearest neighbors and decays rapidly with distance [\figpanel{fig:design}{b}]. At the reference mean coupler--qubit gap of \SI{0.52}{\giga\hertz} used in \figpanels{fig:design}{b}{d}, the median $|g_\mathrm{qq}|/2\pi$ is \SI{3.8}{\mega\hertz} for nearest neighbors but only \SI{17}{\kilo\hertz} at Manhattan distance 2. Tuning the couplers towards the qubits increases the nearest-neighbor coupling by almost three orders of magnitude, from \SI{32}{\kilo\hertz} at a gap of \SI{1.8}{\giga\hertz} to \SI{22}{\mega\hertz} at zero gap [\figpanel{fig:design}{e}]. 

Alternatively, starting again from $H^\mathrm{ec}$, we can remove the earlier constraint $V_{ij} = 0$ and $V_{ijkl} = 0$ between qubit modes in \eqref{eq:eta_coefficients} to eliminate all exchange interactions across the entire system. This alternative flow yields a (approximate) diagonal Hamiltonian:
\begin{equation}
    \Tilde{H} = \sum_{i} \Tilde{H}_{i} \Tilde{n}_i + \sum_{ij} \Tilde{H}_{ij} \Tilde{n}_i \Tilde{n}_j + \mathcal{O}(\Tilde{a}^4, \Tilde{a}^{\dag 3} \Tilde{a}^3),
\label{eq:diagonal_hamiltonian}
\end{equation}
where $\Tilde{n}_i = \Tilde{a}_i^\dag \Tilde{a}_i$ is the excitation-number operator of the diagonal eigenbasis. The two-body coefficients for the diagonal Hamiltonian directly give the density-density interaction $\zeta_{ij}$ between qubits $i$ and $j$:
\begin{equation}
\zeta_{ij} = \Tilde{H}_{ij} + \Tilde{H}_{ji} \quad (i \neq j).
\end{equation}
This interaction is commonly referred to as a ZZ or cross-Kerr coupling for superconducting qubits, and cause the transition energy of qubit $i$ to be conditional on the state of qubit $j$, and vice versa~\cite{pettersonforsComprehensiveExplanationZZ2024}. The induced correlations are a source for coherent phase errors during idling or single-qubit operations, but can equally be exploited to generate two-qubit entangling gates~\cite{dicarloDemonstrationTwoqubitAlgorithms2009, sungRealizationHighFidelityCZ2021}. Like the effective couplings, the ZZ couplings $\zeta_{ij}$ between qubits are typically largest for nearest-neighbor pairs [\figpanel{fig:design}{c}]. At the reference gap, the median $|\zeta_\mathrm{qq}|/2\pi$ is \SI{3.0}{\kilo\hertz} for nearest neighbors, with a maximum of \SI{1.6}{\mega\hertz}, and at Manhattan distance 2 the median is more than three orders of magnitude smaller. As the gap closes, the nearest-neighbor ZZ coupling grows from \SI{0.1}{\kilo\hertz} at a gap of \SI{1.8}{\giga\hertz} to \SI{68}{\kilo\hertz} at zero gap [\figpanel{fig:design}{f}].

Furthermore, tracking the spatial support of the creation operator $a_i^\dagger$ throughout the diagonalization gives a direct measure of eigenstate locality. We define the locality of single-excitation eigenstates as the overlap $L_{ij} = \abs{\! \braket*{1_i}{\Tilde{1}_j} }$ between the state $\bra{1_i} = \bra{0} a_i$, with a single excitation at site $i$ in the original basis, and the eigenstate $\ket{\Tilde{1}_j} = \Tilde{a}_j^\dagger \ket{\Tilde{0}}$ labeled $j$. When the non-excitation-conserving couplings are negligible, it is given by the linear coefficients of the flowed creation operator, $a_i^\dag = \sum_m \alpha_{im} \Tilde{a}_m^\dag + \dots$, as $L_{ij} = |\alpha_{ij}|$; otherwise, it also contains contributions from the pairs of excitations that these couplings admix into the vacuum (Methods; \secref{sec:locality}). At the reference gap, the eigenstates remain largely on their own qubit, with a median $|L_{ii}|^2 = 0.98$ and median weights of $3 \times 10^{-3}$ on the adjacent couplers and only $2 \times 10^{-6}$ on neighboring qubits [\figpanel{fig:design}{d}]. Over the entire sweep, the median weight on neighboring qubits stays below $4 \times 10^{-5}$ and is at least 30 times smaller than the median weight on the adjacent couplers, which indicates that the hybridization is mediated by the couplers rather than by direct qubit--qubit overlap [\figpanel{fig:design}{g}].

\section{Quench dynamics of analog simulators}
\label{sec:analog}

Beyond the design metrics of \secref{sec:design}, the effective models of a device raise a broader question: which observables of a quantum processor can be predicted from them? Analog quantum simulators make this question pressing. They now follow the coherent dynamics of hundreds of interacting qubits~\cite{bohnetQuantumSpinDynamics2016, andersenThermalizationCriticalityAnalogue2025a, manovitzQuantumCoarseningCollective2025}, far beyond the reach of exact classical methods. Hence knowing which of their observables remain classically predictable is central both to verifying these experiments and to delimiting claims of quantum advantage. In superconducting processors, couplers and non-excitation-conserving terms generate longer-range hopping, ZZ couplings, and doublon- (second-excited state; $\ket{2_i}$) mediated interactions, whose magnitudes are set by the full device rather than by design. Predicting the dynamics of such a device therefore requires deriving its effective Hamiltonian from the device model. 

\begin{figure*}
    \centering
    \includegraphics[width=\linewidth]{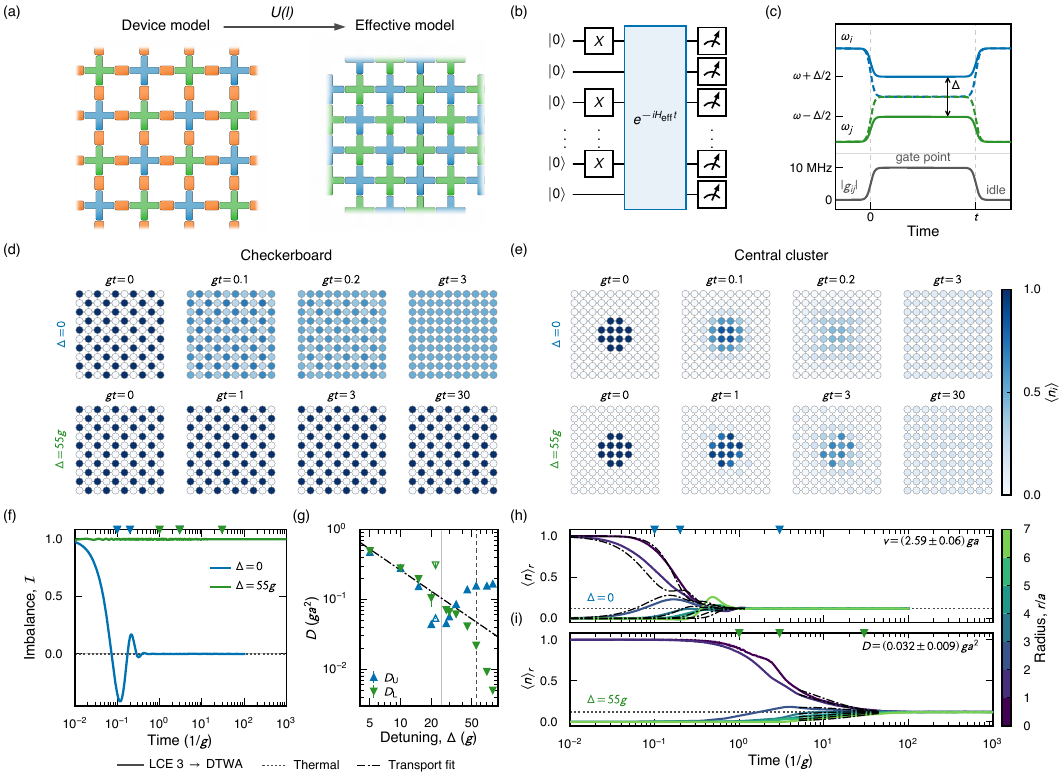}
    \caption{Quench dynamics of a 280-transmon analog simulator, predicted from its device model. 
    (a) A CUT $U(l)$ maps the device model of 100 qubits, on an upper (blue) and a lower (green) sublattice in frequency, and 180 couplers (orange) onto an effective qubit model. 
    (b) Protocol: $X$ gates prepare a Fock product state, the array evolves under the effective Hamiltonian $H_\mathrm{eff}$ for a time $t$, and all qubits are measured. 
    (c) During the evolution, the two qubit sublattices are tuned to $\omega \pm \Delta/2$ and the couplers from the idle point ($g_{ij} = 0$) to the gate point ($|g_{ij}|/2\pi = \SI{10}{\mega\hertz}$). 
    (d) Snapshots of the qubit occupations $\langle n_i \rangle$ after preparing a checkerboard state, at resonance ($\Delta = 0$, top) and for a staggered detuning $\Delta = 55g$ (bottom). 
    (e) As (d) for a central cluster of excitations. 
    (f) Imbalance between the initially occupied and empty sublattices; triangles mark the snapshot times in (d). 
    (g) Diffusion constants of the cluster excitations on the lower ($D_\mathrm{L}$, green down triangles) and upper ($D_\mathrm{U}$, blue up triangles) sublattice versus $\Delta$, with the second-order law $D \propto g^2/\Delta$ (dash-dotted); vertical lines mark the doublon resonance $\Delta = |\alpha_\mathrm{U}| = 25g$ (gray) and the detuning of (i) (dashed); open markers indicate where the second-order hard-core projection of the DTWA is unreliable (Supplementary Information; \figref{fig:ed_detuning_sweep}). 
    (h), (i) Mean occupation on shells of radius $r$ (lattice constant $a$) around the cluster center for $\Delta = 0$ and $\Delta = 55g$, respectively, with fits of ballistic, $\langle r^2 \rangle - \langle r^2 \rangle_0 \approx v^2 t^2$ (h), and diffusive, $\langle r^2 \rangle - \langle r^2 \rangle_0 \approx 4Dt$ (i), transport (dash-dotted) giving the velocity $v$ and the diffusion constant $D$ (Methods; \secref{sec:transport_fits}), where $\langle r^2 \rangle$ is the occupation-weighted mean-square distance from the center of the array. 
    Solid lines: excitation-cluster expansion of order 3 joined to DTWA (LCE3$\,\to\,$DTWA); dotted lines: thermal values.}
    \label{fig:analog_simulation}
\end{figure*}

Here, we establish what CUT makes classically accessible in quench dynamics, and what it does not. Quenches from Fock product states and read out site by site, are the standard probe of thermalization and transport in analog simulators~\cite{schreiberObservationManybodyLocalization2015, andersenThermalizationCriticalityAnalogue2025a}. They also test our approach where it is both demanding and verifiable: when the qubits are resonant, the eigenstates delocalize across the array, whereas the conserved number of excitations provides two references, exact dynamics of small arrays and thermal values at every size. Applied to the architecture of \secref{sec:design} scaled to a device of 280 transmons, CUT yields an effective qubit model that, combined with a semiclassical sampler, predicts relaxation, thermalization, and transport after such a quench with 50 excitations, a sector of dimension $10^{29}$ [\figpanels{fig:analog_simulation}{a}{c}]. The bitstring statistics that underlie cross-entropy benchmarks, by contrast, remain out of reach of our approach.

Of the two effective models of \secref{sec:design}, the diagonal model of \eqref{eq:diagonal_hamiltonian} offers the most direct route from CUT to dynamics, the diagonal route, since time evolution is straightforward in its eigenbasis. Thomson and Eisert used it to obtain infinite-temperature correlators of disordered fermions, which can be sampled over product eigenstates without transforming any initial state~\cite{thomsonUnravellingQuantumDynamics2024}. A quench experiment instead starts from a specific Fock state of the device, which first must be mapped onto the eigenbasis by applying the flowed creation operator of \secref{sec:design} once per excitation. For transmon arrays, this mapping fails when the eigenstates delocalize, i.e., at weak disorder. With an anharmonicity $|\alpha|$ much larger than the exchange coupling $g$, transmons behave as hard-core bosons, but bosonic operators, unlike fermionic ones, do not enforce exclusion: the linear part of the flowed operator places two excitations partly in the same eigenmode, and the non-linear terms that must remove this weight become as large as the linear ones at weak disorder (Supplementary Information; \secref{sec:si_hardcore}). Near resonance, this mapping consequently yields late-time density errors of order unity, which decrease appreciably only for detunings beyond the anharmonicity (Supplementary Information; \figref{fig:ed_renormalizability}), consistent with an error estimate adapted from Ref.~\cite{thomsonUnravellingQuantumDynamics2024} (Supplementary Information; \secref{sec:si_error}). The diagonal route thus fails precisely for disorder weaker than the anharmonicity, the regime in which analog simulators are operated to observe fast thermalization.

We therefore use CUT only for block diagonalization and leave the dynamics to a sampling method. Eliminating first the non-excitation-conserving terms and then the couplers yields the effective qubit model of \eqref{eq:qubit_hamiltonian}, which we compute on overlapping $5 \times 5$ windows for the largest arrays (Methods; \secref{sec:windowed}). It retains long-range hopping (up to \SI{0.35}{\mega\hertz}), ZZ couplings (up to \SI{90}{\kilo\hertz}) and doublon-mediated terms of 8--\SI{12}{\percent} of $g$. As in experiments, we calibrate the bare frequencies and couplings through this model such that every effective nearest-neighbor coupling at the gate point is $|g|/2\pi = \SI{10}{\mega\hertz}$ [\figpanel{fig:analog_simulation}{c}]. At early times, we evolve the model with an excitation-cluster expansion that treats every subset of up to three initial excitations exactly, and we hand over to the discrete truncated Wigner approximation (DTWA)~\cite{polkovnikovPhaseSpaceRepresentation2010, schachenmayerManyBodyQuantumSpin2015} at late times (Methods; \secref{sec:dynamics}).

We quantify errors by the misplaced fraction $\sum_i |n_i - n_i^\mathrm{ref}| / 2n^\mathrm{tot}$, the fraction of the $n^\mathrm{tot}$ excitations that must move to other sites to reproduce reference densities $n_i^\mathrm{ref}$. For a $4 \times 4$ array, the effective model reproduces the exact dynamics of a central cluster in the full array, including couplers, to a misplaced fraction of $0.003$. The error is thus dominated by the sampling: against exact dynamics of the effective model, the misplaced fraction of the sampled densities stays at or below 0.05 for the checkerboard and reaches 0.13 during the early transient of a central cluster. Eliminating the couplers before sampling makes the DTWA both faster and more accurate. For the 16-mode effective model, it runs 13--43 times faster than for the 40-mode array, whose couplers precess about \SI{1}{\giga\hertz} above the qubits. Treating the couplers explicitly gains no accuracy at resonance and loses it where transport rests on weak second-order hopping: at $\Delta = 55g$, the DTWA of the full array underestimates the diffusion constant of the central cluster by a factor of two, predicting $(0.026 \pm 0.002)\,ga^2$ against $(0.054 \pm 0.008)\,ga^2$ in exact dynamics, whereas our approach predicts $(0.051 \pm 0.003)\,ga^2$ from the effective model (Supplementary Information; \secref{sec:si_fullarray_dtwa}) in units of the lattice constant $a$.

At resonance ($\Delta = 0$), the checkerboard imbalance relaxes to its thermal value within $gt \approx 0.35$ after a single overshoot [\figpanel{fig:analog_simulation}{d,f}]. For a staggered detuning $\Delta = 55g$, the imbalance instead stays near unity up to $gt = 10^3$. First-order hopping between the sublattices is then off-resonant, and second-order hopping via the other sublattice, with amplitude $J_2 \sim g^2/\Delta$, connects only sites of the same sublattice, which the checkerboard fills completely. This frozen imbalance reflects equilibrium rather than localization: the occupied sublattice is the one of lower frequency, and the checkerboard lies only $3g$ above the dressed ground state of its excitation sector, at a temperature $T \approx \Delta/10$ where the thermal imbalance is $0.995$ (Supplementary Information; \secref{sec:si_temperature}).

Transport exposes the device-level physics directly. At resonance, a central cluster spreads ballistically, with velocity $v = (2.59 \pm 0.06)\,ga$, until it reaches the edge of the array [\figpanel{fig:analog_simulation}{e,h}]. For $\Delta \geq 5g$, it instead spreads diffusively through $J_2$, with $D = (0.032 \pm 0.009)\,ga^2$ at $\Delta = 55g$ for both sublattices together [\figpanel{fig:analog_simulation}{i}] and $D \propto g^2/\Delta$ on both sublattices up to $\Delta \approx 15g$ [\figpanel{fig:analog_simulation}{g}]. Near the doublon resonance of the upper sublattice, $\Delta = |\alpha_\mathrm{U}| = 25g$, the doublons that defeat the diagonal route reappear: transport on that sublattice is suppressed two- to threefold. Above $\Delta \approx 40g$, the sublattices separate: $D_\mathrm{U}$ saturates at about $0.16\,ga^2$, whereas $D_\mathrm{L}$ falls below $0.01\,ga^2$ at $70g$. This splitting is consistent with a coupler-mediated next-nearest-neighbor hopping of ${\sim}0.03g$, independent of $\Delta$ and absent from the designed model, which adds to $J_2$ on the upper sublattice and opposes it on the lower one. The sublattice-resolved diffusion constants of the $8 \times 8$ and $10 \times 10$ arrays agree to within about \SI{20}{\percent} (Supplementary Information; \figref{fig:ed_detuning_sweep}).

As a semiclassical method, the DTWA reproduces the relaxation and the thermal plateau, but misses the weak quantum fluctuations of the exact dynamics around the plateau (Methods; \figref{fig:lce_dtwa_validation_resonant}), although these decrease as the excitation sector grows. This limitation also keeps bitstring statistics out of reach of our approach: once the bitstring distribution approaches the Porter--Thomas distribution, the cross-entropy of bitstrings sampled from the trajectories falls to about $0.01$ on a $4 \times 4$ array, whereas the experiment of Ref.~\cite{andersenThermalizationCriticalityAnalogue2025a} reaches fidelities of order unity (Supplementary Information; \secref{sec:si_xeb}). Our results therefore do not challenge cross-entropy-based claims of quantum advantage. The approach extends to initial states with few-body entanglement, such as the dimer states of Ref.~\cite{andersenThermalizationCriticalityAnalogue2025a}, through a generalized DTWA~\cite{zhuGeneralizedPhaseSpace2019}, but likely not to extensive multipartite entanglement. 

\section{Conclusion}
We have shown that continuous unitary transformations can turn the device model of a superconducting quantum processor into quantitative predictions at the scale of the processor itself. The decisive advantage of CUTs is that they, at a polynomial cost in qubit numbers and at a saturated amount of ${\sim}10^3$ integration steps (Methods; \secref{sec:scaling}), approximately eliminate the non-excitation-conserving terms and the couplers of two-dimensional transmon arrays, neglecting only the higher-order couplings that the flow generates. This flow leaves an effective qubit model that retains effective qubit couplings, state-dependent interactions (e.g., ZZ couplings), and eigenstate locality useful in processor design. Combined with semiclassical DTWA sampling, this model predicts the relaxation, thermalization, and transport of 280 transmons within tens of GPU-hours, though not the bitstring statistics behind cross-entropy benchmarks. The predictions carry fingerprints of the device that a detuning sweep can experimentally test.

Looking ahead, such models can accompany a processor from layout to operation. Because a calibrated model of 280 transmons only takes hours on a single GPU and a local layout change only affects the windows that contain it (Methods; \secref{sec:windowed}), a candidate layout can be evaluated within a day rather than after a month-long fabrication and calibration cycles. In operation, the same comparison applies to any measurement of densities or transport on the full device, where a deviation beyond the sampling error would point to physics missing from the device model. Including decoherence and time-dependent control in the model would bring gates within reach, and since the reach of the flow is set by the memory of its tensors rather than by entanglement, harnessing tensor symmetries (Methods; \secref{sec:scaling}) and spatial linked-cluster expansions will extend it to devices with thousands of qubits. As processors outgrow exact simulation, CUT-based models cannot stand in for the hardware itself, but they offer a quantitative guide to designing it and verifying its operation.

\section{Methods}

\subsection{Scaling and computational complexity}
\label{sec:scaling}

\begin{figure*}
    \centering
    \includegraphics[width=\linewidth]{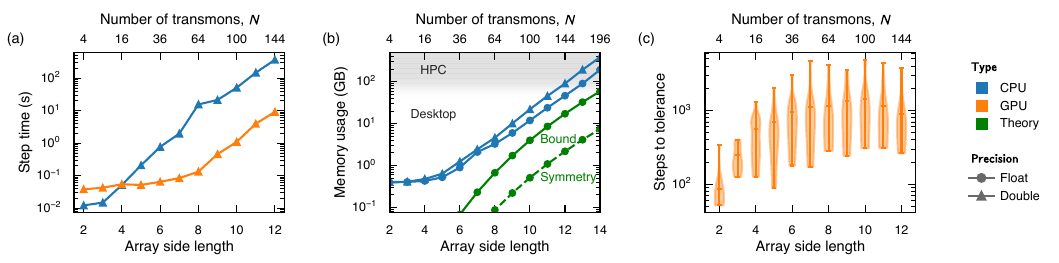}
    \caption{Performance characteristics for the CUT implementation as a function of system size. The characteristics are generated for diagonalization of a square grid of transmons without couplers and non-excitation-conserving couplings.
    (a) Time cost per iteration of the flow-equation solver. 
    (b) Spatial (memory) cost compared to theoretical estimates (Theory; green). 
    (c) Distribution over sampled system realizations of the number of steps to reach a set tolerance condition, indicating convergence. Colors distinguish the hardware (CPU, GPU) from the theoretical estimates, and markers distinguish the data types. The gray area in (b) indicates the approximate crossover from desktop devices to where high-performance computing (HPC) clusters are needed to meet the memory requirements. The solid curve labeled 'Bound' represents the theoretical lowest memory bound for the current implementation, while the dashed curve labeled 'Symmetry' additionally exploits symmetries in the Hamiltonian tensors. (c) is sampled from 21 realizations of system parameters per system size, using the tolerance condition $\mathrm{max}(H_{ij}, H_{ijkl}) / 2\pi < \SI{100}{\kilo\hertz}$ for the off-diagonal elements. The distributions are represented with violin plots, where the mid marks give the means.}
    \label{fig:benchmark}
\end{figure*}

To evaluate the practical applicability and scaling limits of our CUT implementation, we benchmark its computational performance as a function of system size $N$ (\figref{fig:benchmark}). Both computation time per ordinary differential equation (ODE) integration step and memory consumption scale polynomially with $N$ [\figpanels{fig:benchmark}{a}{b}]. Leveraging GPU acceleration yields a one-to-two order-of-magnitude reduction in step execution time compared to CPU alternatives, owing to the high degree of parallelizability inherent to tensor contractions in the flow equations. Consequently, diagonalizing arrays of 64 transmons requires only minutes of total compute time, whereas systems exceeding 100 transmons complete within single-digit hours. Because step evaluation times scale favorably on parallel hardware, spatial memory availability emerges as the primary operational constraint [\figpanel{fig:benchmark}{b}]. On single-node desktop workstations ($\sim \SI{64}{\giga\byte}$ RAM), CUT-based diagonalization is feasible up to $N \approx 120$ transmons, whereas high-performance computing (HPC) CPU clusters extend this reach beyond $N = 200$.

A critical factor governing total runtime is the number of ODE integration steps required for the flow to reach target convergence. If step counts scaled aggressively with system size, numerical integration would impose a severe computational barrier. However, sampled distributions across physical parameter realizations demonstrate that the required integration steps remain remarkably flat as $N$ grows [\figpanel{fig:benchmark}{c}]. The median step count increases modestly for small arrays before saturating to approximately $10^3$ steps near $N = 49$. We attribute this saturation to a boundary-bulk crossover: as array dimensions expand, the fraction of boundary sites scales as $\mathcal{O}(N^{-1/2})$, causing the renormalization flow in 2D grids to become rapidly dominated by uniform bulk transmon environments. Furthermore, the broad variance in step distributions indicates that convergence rates depend far more strongly on local parameter detunings and microscopic disorder than on overall array size.

To extend CUT simulations to even larger architectures, we examine spatial optimization strategies [\figpanel{fig:benchmark}{b}]. The theoretical lower bound on memory is dictated by the maximum number of intermediate tensors that must be held simultaneously during ODE integration. For a standard explicit Runge–Kutta 45 (RK45) scheme, this working set comprises ten pairs of rank-2 and rank-4 coefficient tensors (representing the evolving Hamiltonian, generator, and internal solver stages). Streamlining memory allocations to this strict operational bound [Fig.~\ref{fig:benchmark}(b), solid green curve] provides a factor-of-two memory reduction for $N \ge 10$, showing that current memory usage is close to its algorithmic limit. Alternative integration schemes, such as lower-order RK23 adaptive step solvers, could further reduce spatial memory footprint at the expense of additional integration steps.

A far more substantial memory reduction is theoretically accessible by exploiting internal symmetries inherent to the operator representations (Table~\ref{tab:cost_scaling}). For two-body interactions ($K=2$), Hermiticity together with the commutation relations of bosonic operators ($[a_i^\dagger, a_j^\dagger] = [a_k, a_l] = 0$) imposes that parameterizations are non-unique and can be chosen fully symmetric under index swaps ($i \leftrightarrow j$ and $k \leftrightarrow l$). These intrinsic symmetries reveal that, for large $N$, $H_{ij}$ and $H_{ijkl}$ contain only $N^2/2$ and $N^4/8$ independent elements, respectively—yielding a spatial compression factor of $1/8$ for two-body interaction tensors ($K=2$ in Table~\ref{tab:cost_scaling}). Projected estimates show that symmetry-adapted tensor representations would lower memory requirements by nearly an order of magnitude [Fig.~\ref{fig:benchmark}(b), green dashed curve], theoretically extending desktop-level diagonalization capacity beyond $N \approx 150$ transmons. While exploiting these symmetries in software will require custom tensor-contraction routines to avoid expanding back to dense arrays during integration, this scaling estimate highlights a clear and practical pathway for future algorithmic development to reach multi-hundred-qubit architectures.

\begin{table}
    \centering
        \caption{Memory footprint and leading computational contraction costs per flow step, listed for full and symmetry-reduced parameterizations of order $2K$ with system dimension $N$, evaluated in the large-$N$ limit. The symmetry reduction factor denotes the ratio of unique tensor elements (memory) or required floating-point operations (computation) relative to the unconstrained dense tensor format. For $K=2$, memory and operation counts are both compressed by a factor of $1/8$, directly corresponding to the memory savings shown in \figpanel{fig:benchmark}{b}.}
        \label{tab:cost_scaling}
        \small
        \setlength{\tabcolsep}{3.5pt}
        \renewcommand{\arraystretch}{1.25}
        \begin{tabular}{l c c c c}
        \toprule
        $K$ & \shortstack{Memory \\(Full)} & \shortstack{Memory \\Reduction} & \shortstack{Computation \\(Full)} & \shortstack{Computation \\Reduction} \\
        \midrule
        $2$ & $N^4$ & $1/8$ & $N^6$ & $1/8$ \\
        $3$ & $N^6$ & $1/72$ & $N^9$ & $1/216$ \\
        $4$ & $N^8$ & $1/1152$ & $N^{12}$ & $1/13\,824$ \\
        $5$ & $N^{10}$ & $1/28\,800$ & $N^{15}$ & $1/1\,728\,000$ \\[2pt]
        \midrule
        $K$ & $N^{2K}$ & $\dfrac{1}{2(K!)^2}$ & $N^{3K}$ & $\dfrac{1}{(K!)^3}$ \\
        \bottomrule
        \end{tabular}
    \end{table}

Beyond system size $N$, systematic refinement of the CUT expansion requires incorporating higher-order $K$-body interaction terms. Table~\ref{tab:cost_scaling} summarizes the asymptotic memory and computational complexity, alongside symmetry-reduction factors, as a function of truncation order $K$. While spatial memory scales as $\mathcal{O}(N^{2K})$, the computational cost per flow step grows as $\mathcal{O}(N^{3K})$. Consequently, for higher $K$, execution time, rather than memory, becomes the dominant bottleneck. Crucially, exploiting tensor symmetries mitigates this computational burden even more efficiently than memory usage: symmetry reduction compresses computational operation counts by $1/(K!)^3$, compared to $1/[2(K!)^2]$ for memory. Without symmetry compression, extending CUT expansions beyond $K=2$ becomes computationally prohibitive for intermediate systems ($N \sim 25\text{--}50$).

Finally, the scaling behavior summarized in \tabref{tab:cost_scaling} clarifies the cost relationship between CUTs and exact diagonalization (ED) in excitation-conserving systems. A sector of $K$ excitations has dimension $\mathcal{O}(N^K)$, and its complete diagonalization requires $\mathcal{O}(N^{2K})$ storage and $\mathcal{O}(N^{3K})$ operations. A flow truncated at $K$-body terms has the same asymptotic complexity per step, and its number of steps saturates with $N$ [\figpanel{fig:benchmark}{c}]. With normal ordering with respect to the vacuum, this flow is moreover exact in all sectors of up to $K$ excitations, and its coefficients, including those of the flowed operators, are fixed by these sectors alone, since couplings of more bodies never feed back into couplings of fewer (Supplementary Information; \secref{SI:unidirectional}). 

Consequently, CUTs do not provide an asymptotic speedup over ED for complete sector diagonalization with vacuum normal ordering, and standard linear-algebra routines carry less overhead than ODE solvers. This includes the approximate diagonalization of disordered fermions in Ref.~\cite{thomsonUnravellingQuantumDynamics2024}, whose diagonal Hamiltonian and flowed operators follow from the exact one- and two-particle sectors. ED alone does not, however, supply the labeling: the flow connects each eigenstate continuously to a bare Fock state, and thereby each two-excitation eigenstate to a pair of single-excitation modes, whereas ED requires a separate matching step \cite{pettersonforsComprehensiveExplanationZZ2024}, for instance by maximal overlap with products of single-excitation eigenstates, which becomes ambiguous near resonances. The same equivalence extends to the block diagonalization of excitation-conserving Hamiltonians, which ED of these sectors followed by a direct rotation within each of them~\cite{bravyiSchriefferWolffTransformationQuantum2011} achieves up to the labeling of the blocks and the choice of frame. The decisive advantage of CUTs lies instead in the block diagonalization of non-excitation-conserving Hamiltonians. Since these couplings connect all excitation sectors, matrix methods such as the exact Schrieffer--Wolff transformation~\cite{bravyiSchriefferWolffTransformationQuantum2011} require the eigenprojectors of the full Hamiltonian, at a cost exponential in $N$, whereas CUTs eliminate the couplings across all sectors simultaneously while retaining a compact polynomial operator representation.

\subsection{Simulation details}
\label{sec:computing}

\subsubsection{Flow equations}

All commutators are evaluated as contractions of the coefficient tensors, implemented in JAX~\cite{bradburyJAXComposableTransformations2018}. The flow equation is integrated with the adaptive fifth-order Runge--Kutta method Tsit5 of the Diffrax library~\cite{kidgerNeuralDifferentialEquations2021}, with relative and absolute tolerances of $10^{-6}$ and $10^{-8}$, and is stopped once the largest off-diagonal coupling falls below a set threshold. The thresholds depend on the particular problem at hand, but are in general chosen to be consistent with the approximation of neglecting higher-order interactions, e.g., three-body couplings, and informed by \figref{fig:nec_validation}. All production calculations use double-precision arithmetic.

\subsubsection{Design sweep}
\label{sec:design_sweep}

The device of \figref{fig:design} has qubit frequencies drawn around 4.0 and \SI{4.75}{\giga\hertz} on the two sublattices, qubit anharmonicities of $-150$ and \SI{-250}{\mega\hertz}, coupler anharmonicities of $-150$ to \SI{-225}{\mega\hertz} in four staggered groups, and couplings of \SI{50}{\mega\hertz} (qubit--coupler), \SI{1.3}{\mega\hertz} (qubit--qubit) and \SI{5}{\mega\hertz} (coupler--coupler), all with a relative spread of \SI{5}{\percent}. Qubit frequencies closer than \SI{0.1}{\giga\hertz} are then pushed apart, which widens the band to 2.3--\SI{6.5}{\giga\hertz}. Each coupler $k$ between qubits $i$ and $j$ starts at the frequency above both qubits at which their second-order coupling, including the non-excitation-conserving terms, vanishes, i.e.,
\begin{equation}
    H_{ij} + H_{ik} H_{jk} \sum_{m = i, j} \frac{\omega_k}{\omega_m^2 - \omega_k^2} = 0 ,
\end{equation}
with $\omega_i$ the bare frequencies and $H_{ij}$ the bare couplings of \eqref{eq:device_hamiltonian}. This condition places the couplers at 5.3--\SI{8.2}{\giga\hertz}. Each coupler is then tuned linearly towards the higher of its two qubit frequencies, all couplers through the same fraction of their range, in 25 equidistant steps that take the mean gap from \SI{1.79}{\giga\hertz} to zero. At each step, the non-excitation-conserving terms are eliminated once, and two flows start from the resulting $H^\mathrm{ec}$: the elimination of the couplers, which yields $g_{ij}$, and the full diagonalization, which yields $\zeta_{ij}$ and, through the flowed creation operator, $L_{ij}$. The thresholds are \SI{10}{\kilo\hertz} for the non-excitation-conserving flow and \SI{100}{\kilo\hertz} for the diagonalizing flow, and, for the flow to the qubit model, \SI{10}{\hertz} on $V_{ij}$ and \SI{1}{\mega\hertz} on $V_{ijkl}$. All results are for a single realization of the device parameters.

\subsubsection{Windowed flow}
\label{sec:windowed}

For the arrays of \figref{fig:analog_simulation}, we run the flow on overlapping $5 \times 5$ windows of the bare model rather than on the full array, exploiting the rapid decay of the effective couplings with distance. We use the system parameters and solver settings in \tabref{tab:numerics}. Each window keeps its qubits, the couplers between them and the dangling couplers to neighboring qubits, so that every qubit retains its coupler environment. Windows overlap such that every pair of qubits at most two sites apart lies at least one site from an artificial window edge in some window, and each coefficient of the effective model is taken from the window in which its qubits lie deepest. Coefficients that fit into no window are set to zero; from the decay of the window couplings, the largest dropped coupling is estimated at \SI{0.03}{\kilo\hertz}. For the $6 \times 6$ array with couplers, windowed and full-array flows at identical bare parameters agree to within \SI{0.06}{\kilo\hertz} in the frequencies and \SI{0.014}{\kilo\hertz} in the hopping. The bare frequencies and couplings are calibrated iteratively through the windowed model until all effective frequencies and nearest-neighbor couplings are within \SI{1}{\kilo\hertz} of their targets. Each window has a fixed size and takes about \SI{100}{\second} per calibration iteration on a NVIDIA H100 GPU, so the cost of the effective model grows linearly with the number of windows, $(L-4)^2$ for an $L \times L$ array, and a local change of the device only affects the windows that contain it.

\begin{table}
\centering
    \caption{Numerical parameters of the quench dynamics in \figref{fig:analog_simulation}. Frequencies and couplings are given as $\omega/2\pi$, etc., bare couplings as nominal values, and the four coupler anharmonicities are spaced by \SI{25}{\mega\hertz}.}
    \label{tab:numerics}
    \small
    \setlength{\tabcolsep}{3.5pt}
    \renewcommand{\arraystretch}{1.25}
    \begin{tabular}{l c}
    \toprule
        Parameter & Value \\
    \midrule
        Target frequency, lower qubit sublattice & \SI{4.0}{\giga\hertz} \\
        Qubit anharmonicities, lower, upper & $-150$, \SI{-250}{\mega\hertz} \\
        Coupler anharmonicities & $-150$ to \SI{-225}{\mega\hertz} \\
        Qubit--coupler coupling & \SI{150}{\mega\hertz} \\
        Direct qubit--qubit coupling & \SI{7}{\mega\hertz} \\
        Direct coupler--coupler coupling & \SI{5}{\mega\hertz} \\
        Spread of couplings to couplers & \SI{1}{\percent} \\
    \midrule
        Non-excitation-conserving threshold & \SI{1}{\kilo\hertz} \\
        Coupler-flow threshold, $V_{ij}$ & \SI{1}{\kilo\hertz} \\
        Coupler-flow threshold, $V_{ijkl}$ & \SI{100}{\kilo\hertz} \\
        Maximum flow time, stages 1, 2 & 50, 5000 \\
        Window side, buffer, reach & 5, 1, 2 \\
        Windows ($10 \times 10$ array) & 36 \\
        Calibration tolerance & \SI{1}{\kilo\hertz} \\
    \midrule
        Excitation-cluster order & 3 \\
        Tensor threshold (from 64 qubits) & \SI{e-7}{\giga\hertz} \\
        DTWA samples & 20\,000 \\
        DTWA output step & $2.5 \times 10^{-3}/g$ \\
        RK4 substeps ($\Delta = 0$, $55g$) & 2, 9 \\
        Maximum $gt$ ($\Delta = 0$, $55g$) & 100, 1000 \\
    \bottomrule
    \end{tabular}
\end{table}

\subsubsection{Dynamics}
\label{sec:dynamics}

The effective model is evolved in two complementary ways. The first is an expansion in clusters of initial excitations. It follows numerical linked-cluster expansions~\cite{rigolNumericalLinkedClusterApproach2006, tangShortIntroductionNumerical2013, mallayyaQuantumQuenchesRelaxation2018}, but its clusters are subsets of excitations rather than of lattice sites, as in the cluster-correlation expansion of spin-bath decoherence~\cite{yangQuantumManybodyTheory2008}, and each cluster evolves on the full array. For a Fock product state with excitations on the set $\mathcal{O}$ of initially occupied qubits, $|\mathcal{O}| = n^\mathrm{tot}$, let
\begin{equation}
    n_i^S(t) = \bra{S} \e^{\i H' t}\, a_i^\dag a_i\, \e^{-\i H' t} \ket{S}, \quad \ket{S} = \prod_{j \in S} a_j^\dag \ket{0},
\end{equation}
be the density of qubit $i$ when only the excitations in $S \subseteq \mathcal{O}$ are present, with $n_i^\emptyset = 0$. As the effective model conserves the number of excitations, $n_i^S$ is computed exactly in the sector of $|S|$ excitations. The connected contributions $\kappa_i^S$, defined by $n_i^S = \sum_{T \subseteq S} \kappa_i^T$, follow by Möbius inversion over subsets,
\begin{equation}
    \kappa_i^S(t) = \sum_{T \subseteq S} (-1)^{|S| - |T|}\, n_i^T(t),
\end{equation}
and the exact density is $n_i(t) = n_i^\mathcal{O}(t) = \sum_{S \subseteq \mathcal{O}} \kappa_i^S(t)$. The contribution $\kappa_i^S$ vanishes whenever $S$ splits into groups of excitations that evolve independently; for a quadratic model, $n_i^S$ is additive in the excitations and only single excitations contribute. The expansion of order $k$ (LCE$k$) keeps the clusters of up to $k$ excitations,
\begin{equation}
    n_i^{\mathrm{LCE}k}(t) = \sum_{k' = 1}^{k} \delta_{k',i}(t), \quad \delta_{k,i}(t) = \sum_{|S| = k} \kappa_i^S(t),
\end{equation}
and is exact for $k = n^\mathrm{tot}$. Since each subset of $k'$ excitations lies in $\binom{n^\mathrm{tot} - k'}{k - k'}$ clusters of size $k$, the contribution of order $k$ follows from the cluster sums alone,
\begin{equation}
    \delta_{k,i} = \sum_{|S| = k} n_i^S - \sum_{k' = 1}^{k - 1} \binom{n^\mathrm{tot} - k'}{k - k'}\, \delta_{k',i},
\end{equation}
without storing the individual $\kappa_i^S$. The $n_i^S$ are obtained by block Chebyshev propagation in sparse fixed-excitation sectors.

The DTWA~\cite{polkovnikovPhaseSpaceRepresentation2010, schachenmayerManyBodyQuantumSpin2015} operates on the hard-core projection of the effective model, \eqref{eq:hard_core_model}, whose interaction $V_{ij}$ contains the ZZ couplings and the second-order interaction through virtual doublons, supplemented by doublon-assisted correlated hopping; the doublon-formation amplitudes are taken from the two-body tensor of the effective model. Each trajectory starts from a sample of the discrete Wigner distribution of the Fock product state~\cite{schachenmayerManyBodyQuantumSpin2015}: the spin component $s_i^z = \pm 1$ is fixed by the occupation of qubit $i$, and the transverse components $s_i^x, s_i^y = \pm 1$ are drawn independently with equal probability. Each trajectory is integrated with a fourth-order Runge--Kutta method, with substeps chosen to resolve the fastest precession. The two results are joined as $n_i(t) = w(t)\, n_i^\mathrm{LCE3}(t) + [1 - w(t)]\, n_i^\mathrm{DTWA}(t)$, with
\begin{equation}
    w(t) = \frac{1}{2} \mathrm{erfc}\mleft[ \frac{\ln(\sqrt{2}\, t / t^*)}{\sqrt{2}\, \sigma} \mright], \quad \sigma = \frac{\ln 2}{2},
\end{equation}
which falls from 0.84 at $t^*/2$ to 0.16 at $t^*$. Using the misplaced fraction of \secref{sec:analog}, $\mathcal{D}(x) = \sum_i |x_i| / 2n^\mathrm{tot}$ for $n^\mathrm{tot}$ excitations, applied to the contributions $\delta_k = (\delta_{k,i})_i$ of order $k$, the trust time $t^*$ is the first time at which the estimated error $E_3(t) = \mathcal{D}(\delta_3)^2 / \max_{t' \leq t} \mathcal{D}(\delta_2)$ exceeds both 0.05 and the disagreement $\mathcal{D}(n^\mathrm{LCE3} - n^\mathrm{DTWA})$.

\subsubsection{Transport fits}
\label{sec:transport_fits}

The transport coefficients are fitted to the qubit occupations $n_i(t)$ after the central-cluster quench, with the mean-square radius $\langle r^2 \rangle = \sum_i n_i |\mathbf{r}_i - \mathbf{r}_0|^2 / \sum_i n_i$ measured from the center $\mathbf{r}_0$ of the array. At resonance, and for $\Delta \leq 2g$ (Supplementary Information; \figpanel{fig:ed_detuning_sweep}{a}), we fit the lattice telegrapher equation $\tau \ddot{n} + \dot{n} = -D \mathcal{L} n$, with $\mathcal{L}$ the graph Laplacian of the nearest-neighbor bonds, from $t = 0$ with $\dot{n}(0) = 0$. It describes ballistic spreading, $\langle r^2 \rangle - \langle r^2 \rangle_0 = v^2 t^2$ with $v = \sqrt{2D/\tau}$, for $t \ll \tau$ and diffusion for $t \gg \tau$, and its underdamped modes capture the reflection of the front at the edges of the array. The fit ranges end where $\langle r^2 \rangle$ has completed 50, 80, and \SI{98}{\percent} of its growth, the last extended to three times that time; $v$ is taken from the longest range, with an uncertainty of half its spread over the three. For $\Delta \geq 5g$, first-order hopping between the sublattices is off-resonant, and we fit the diffusion equation $\dot{n} = -D \mathcal{L}_2 n$ on the graph of second-order hops within each sublattice, which connects diagonal neighbors (two paths, weight 2) and next-nearest neighbors along rows and columns (one path, weight 1), scaled such that $\langle r^2 \rangle - \langle r^2 \rangle_0 = 4Dt$ in the bulk. The fit starts from the simulated occupations at the beginning of the window in which $\langle r^2 \rangle$ grows from 50 to \SI{95}{\percent} of its final value, and its uncertainty is half the range over two further windows (30--90 and 40--\SI{93}{\percent}). $D$ is fitted to both sublattices together, and $D_\mathrm{L}$ and $D_\mathrm{U}$ to the excitations on the lower and upper sublattice, with $\mathcal{L}_2$ restricted to that sublattice.

\subsubsection{Approximations}

The flow is truncated at normal-ordered strings of at most four ladder operators, retaining the operator content of \eqref{eq:device_hamiltonian}; the strings of four creation or four annihilation operators that the non-excitation-conserving flow generates are discarded as well. For the excitation-conserving flow, whose retained strings are the one- and two-body terms, this truncation leaves the sectors of up to two excitations exact, whereas the non-excitation-conserving flow is controlled by the scale separation $E_{ij} \gg |V^\mathrm{nec}_{ij}|$ (Supplementary Information; \secref{SI:unidirectional}). 

We consider closed systems: decoherence is not included, and all dynamics are the coherent evolution of the device model. In \figref{fig:design} and \figref{fig:analog_simulation}, we assume that the couplers occupy their ground state: the effective qubit model is the block of the flowed Hamiltonian without coupler excitations, and the ZZ couplings and locality overlaps are evaluated with all couplers in their dressed ground state. Thermal or leaked coupler excitations, which would change these coefficients, are neglected. 

The frequency ramps of \figpanel{fig:analog_simulation}{c} enter in two limits. With respect to the qubit--coupler hybridization, we assume them adiabatic, such that they convert the bare qubit states used for preparation and readout into the dressed qubit basis of the effective model, as in the experiment of Ref.~\cite{andersenThermalizationCriticalityAnalogue2025a}, where ramps of more than \SI{5}{\nano\second} suffice. With respect to the qubit dynamics, we treat them as instantaneous: the evolution under the effective model at the gate point starts from the Fock state at $t = 0$, and the hopping and phases accumulated during the ramps are neglected, which requires ramps short compared with the hopping time $1/g \approx \SI{16}{\nano\second}$.

\subsubsection{Hardware and cost}

Production simulations ran on NVIDIA GPUs: H200 GPUs for \figref{fig:design} and \figref{fig:benchmark}, and H100 GPUs for \figref{fig:analog_simulation}. The benchmarks of \figref{fig:benchmark} additionally ran on 64 cores of Intel Xeon Platinum 8358 CPUs, in single and double precision. The four-transmon validations of \secref{sec:sign_validation} ran on a laptop CPU. For the 280 transmons in \figref{fig:analog_simulation}, the effective model takes 5--7 hours on a single GPU and the dynamics at most 6 hours per initial state, except for the checkerboard initial state at $\Delta = 55g$, whose excitation-cluster expansion (LCE3) over all 19\,600 excitation triples of its 50 excitations takes 90 GPU hours, distributed over four GPUs.

\subsection{Eigenstate locality}
\label{sec:locality}

The locality $L_{ij} = \abs{\! \braket*{1_i}{\Tilde{1}_j} }$ of \secref{sec:design} is the overlap of two states defined in different frames: the bare state $\ket{1_i} = a_i^\dag \ket{0}$, with a single excitation at site $i$ on the vacuum $\ket{0}$ of the bare operators in \eqref{eq:device_hamiltonian}, and the single-excitation eigenstate $\ket{\Tilde{1}_j} = \Tilde{a}_j^\dag \ket{\Tilde{0}}$ of the diagonal Hamiltonian in \eqref{eq:diagonal_hamiltonian}. Rather than constructing either state, we transform the bare vacuum and creation operators with the same generators as the Hamiltonian, integrating their coefficient tensors alongside the flow, and evaluate the overlap from the resulting tensors.

The generator of the non-excitation-conserving flow creates and annihilates pairs of excitations. In the $U(1)$-symmetric frame, the bare vacuum therefore acquires a two-excitation component, and the bare creation operator a component of annihilation operators,
\begin{align}
    \ket{0} &= c \ket{0^\mathrm{ec}} + \sum_{ij} c_{ij} \ket{1_i^\mathrm{ec}, 1_j^\mathrm{ec}}, \\
    U_\mathrm{nec} a_i^\dag U_\mathrm{nec}^\dag &= \sum_m \mleft( a_{im} a^{\mathrm{ec} \dag}_m + b_{im} a^\mathrm{ec}_m \mright),
\end{align}
where $\ket{0^\mathrm{ec}}$ is the vacuum of $H^\mathrm{ec}$, which approximates the ground state of the device. We transform these coefficients with the quadratic part of the generator, which truncates the vacuum at two excitations and the creation operator at linear order. The coefficients $c_{ij}$ and $b_{im}$ are of first order in $|V^\mathrm{nec}_{ij}| / E_{ij}$, and the neglected terms are of higher order.

The diagonalizing flow conserves the excitation number. It therefore leaves the vacuum invariant, $\ket{0^\mathrm{ec}} = \ket{\Tilde{0}}$, and maps the creation operators onto
\begin{equation}
    U^\mathrm{ec} a^{\mathrm{ec} \dag}_i U^{\mathrm{ec} \dag} = \sum_{m} \alpha_{im} \Tilde{a}_m^\dag + \sum_{mnl} \alpha_{imnl} \Tilde{a}_m^\dag \Tilde{a}_n^\dag \Tilde{a}_l,
\end{equation}
truncated at three operators, consistent with the two-body truncation of the Hamiltonian. The non-linear coefficients $\alpha_{imnl}$ carry the hard-core constraint of the transmons (Supplementary Information; \secref{sec:si_hardcore}). In the eigenbasis, the two-excitation component of the bare vacuum reads $\sum_{ij} \Bar{c}_{ij} \ket{\Tilde{1}_i, \Tilde{1}_j}$, where $\ket{\Tilde{1}_i, \Tilde{1}_i} = \ket{\Tilde{2}_i}$ denotes the normalized doubly occupied state, such that $\ket{0} = c \ket{\Tilde{0}} + \sum_{ij} \Bar{c}_{ij} \ket{\Tilde{1}_i, \Tilde{1}_j}$.

Writing the locality as $L_{ij} = \abs{\! \mel*{\Tilde{1}_j}{a_i^\dag}{0} }$ and inserting both transformations gives
\begin{equation}
    L_{ij} = \left| c \sum_{m} a_{im} \alpha_{mj} + \sum_{m,n} b_{im} \alpha_{mn} \Bar{c}^{\,s}_{nj} + \sum_{m,n,o} b_{im} \alpha_{mnoj} \Bar{c}^{\,s}_{no} \right|,
    \label{eq:locality_overlap}
\end{equation}
where the symmetrized vacuum coefficient $\Bar{c}^{\,s}_{ij} = (\Bar{c}_{ij} + \Bar{c}_{ji}) / [1 + (\sqrt{2} - 1) \delta_{ij}]$ combines the two orderings of each pair and the normalization of doubly occupied modes. The first term creates the excitation on the vacuum through the linear parts of both transformations. In the other two, the non-excitation-conserving part $b_{im}$ of the bare creation operator instead annihilates an excitation of the pairs in the vacuum. Through the linear coefficients $\alpha_{mn}$, it removes one excitation of a pair and leaves the other in mode $j$, and through the non-linear coefficients $\alpha_{mnoj}$, it converts a whole pair into a single excitation in mode $j$. Both contributions are of second order in $|V^\mathrm{nec}_{ij}| / E_{ij}$. Without non-excitation-conserving couplings, $c = 1$, $c_{ij} = b_{im} = 0$ and $a_{im} = \delta_{im}$, and \eqref{eq:locality_overlap} reduces to $L_{ij} = |\alpha_{ij}|$, as stated in \secref{sec:design}.

\subsection{Sign-generator validation}
\label{sec:sign_validation}

We validate the sign generator in \eqref{eq:eta_coefficients} on a $2 \times 2$ array of four transmons without couplers, for which the device model can be diagonalized exactly. For simplicity, the transmons are modeled as anharmonic oscillators with the Hamiltonian
\begin{equation}
    H = \sum_i \mleft( \omega_i a_i^\dag a_i + \frac{\alpha_i}{2} a_i^\dag a_i^\dag a_i a_i \mright) + g \sum_{\langle ij \rangle} \mleft( a_i^\dag + a_i \mright) \mleft( a_j^\dag + a_j \mright),
    \label{eq:validation_hamiltonian}
\end{equation}
where $\langle ij \rangle$ runs over the four nearest-neighbor pairs. In all comparisons, the exchange coupling is $g/2\pi = \SI{50}{\mega\hertz}$ and the anharmonicities alternate between $\alpha_i/2\pi = \SI{-250}{\mega\hertz}$ and \SI{-150}{\mega\hertz} between nearest-neighbor transmons. Energies are reported in units of $2\pi \times \mathrm{GHz}$, meaning that the flow time $l$ is in units of $(2\pi \times \mathrm{GHz})^{-1}$.

\begin{figure}
    \centering
    \includegraphics[width=\linewidth]{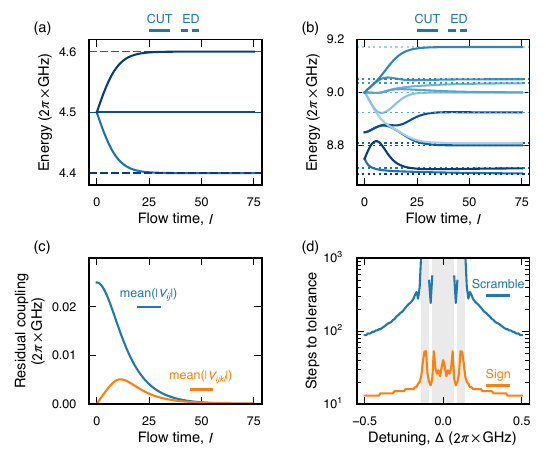}
    \caption{Validation of the sign generator for the excitation-conserving flow of four transmons. 
    (a) Single-excitation and 
    (b) two-excitation energies during the flow of the resonant array (solid), converging to the values from exact diagonalization (ED; thin dashed lines, respectively). 
    (c) Mean magnitude of the residual off-diagonal couplings $V_{ij}$ and $V_{ijkl}$ for the same flow. 
    (d) Accepted solver steps until convergence as a function of the detuning $\Delta$ for the sign generator and for scrambling alternated with Wegner flow. Gray regions mark where the scrambled flows do not converge.}
    \label{fig:validation}
\end{figure}

To test the diagonalizing flow given by \eqref{eq:eta_coefficients}, we retain only the excitation-conserving part of \eqref{eq:validation_hamiltonian} and use four levels per transmon for exact diagonalization (ED). We first consider the resonant array, $\omega_i/2\pi = \SI{4.5}{\giga\hertz}$ for all $i$, where the bare single-excitation energies are degenerate and the Wegner generator vanishes for the nearest-neighbor couplings. The sign generator remains non-vanishing, and the flow reaches the hybridized eigenenergies without external intervention [\figpanels{fig:validation}{a}{c}]. Already at $l = 75$, the eigenvalues of the flowed Hamiltonian agree with ED to within \SI{0.1}{\kilo\hertz} in the one- and two-excitation sectors, while small residual couplings still mix near-degenerate two-excitation states. Continuing the flow until all residual couplings are below \SI{1}{\kilo\hertz}, at $l \approx 860$, the diagonal elements alone reproduce the same energies to within \SI{0.1}{\kilo\hertz}.

We then compare the convergence of the sign generator with that of the scrambling generator~\cite{thomsonUnravellingQuantumDynamics2024}, alternated with Wegner flow, over 101 values of a detuning $\Delta/2\pi$ between $-0.5$ and \SI{0.5}{\giga\hertz}. The frequencies are $\omega_i/2\pi = 4.5\,\mathrm{GHz} + (1, 2.5, -2.5, -1) \times \Delta/2\pi$, with the first and last transmon on the sublattice having the larger anharmonicity (with respect to magnitude). The scrambling transformation is applied whenever a coupling exceeds half the corresponding energy difference, $V_{ij} > \abs{E_i - E_j}/2$, and both flows use the same Runge--Kutta solver and tolerances. Here, we consider that a flow has converged once both maximums of $|V_{ij}|$ and $|V_{ijkl}|$ have both fallen below \SI{100}{\kilo\hertz}. The sign generator converges at all 101 detunings, in at most 53 accepted steps [\figpanel{fig:validation}{d}]. The scrambled flow fails to converge within the integration budget at 21 detunings, in two windows around $\Delta = 0$ and $|\Delta|/2\pi \approx \SI{0.1}{\giga\hertz}$, and elsewhere takes 7--49 times as many steps (median 10).

\begin{figure}
    \centering
    \includegraphics[width=\linewidth]{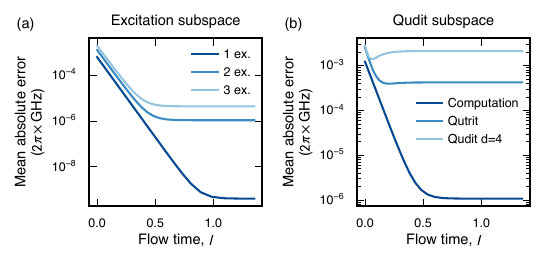}
    \caption{Elimination of the non-excitation-conserving couplings of four non-degenerate transmons with the sign generator. Mean absolute error between the eigenenergies of the full device model and of the excitation-conserving part of the flowed Hamiltonian during the flow, for (a) the sectors of one to three excitations and (b) states grouped by the highest occupation of any transmon, which is one for the computational states (Computation), two for Qutrit and three for Qudit $d = 4$.}
    \label{fig:nec_validation}
\end{figure}

The elimination of the non-excitation-conserving couplings is validated on the same array with the full Hamiltonian in \eqref{eq:validation_hamiltonian}. Here, the frequencies are non-degenerate, $\omega_i/2\pi = (4.0, 4.5, 4.3, 4.1)$\,GHz, ED uses five levels per transmon, and the flow is terminated when the largest non-excitation-conserving coupling falls below \SI{1}{\kilo\hertz}, which occurs at $l \approx 1.4$ after 27 steps. \figref{fig:nec_validation} compares the eigenenergies of the excitation-conserving part of the flowed Hamiltonian with those of the full device model; at $l = 0$, the error is therefore that of simply discarding the non-excitation-conserving terms, i.e., the rotating-wave approximation. For the sectors of one, two, and three excitations, the flow reduces the mean absolute error from 0.6, 1.3, and \SI{1.9}{\mega\hertz} to below \SI{1}{\hertz}, \SI{1.1}{\kilo\hertz}, and \SI{4.4}{\kilo\hertz}, respectively, and from \SI{1.2}{\mega\hertz} to \SI{1.1}{\kilo\hertz} for the computational states. All flows use double-precision arithmetic. The error saturates at higher values for states with higher occupation of a single transmon, at \SI{0.4}{\mega\hertz} when the most occupied transmon holds two excitations and at \SI{2.1}{\mega\hertz} when it holds three, in the latter case close to its initial value. The truncation is thus accurate for the few-excitation physics per transmon considered in this work, whereas more highly excited transmon states require a higher-order operator expansion or a dedicated treatment.

\subsection{Validation of the dynamics}
\label{sec:lce3_dtwa_validation}

\begin{figure*}
    \centering
    \includegraphics[width=\linewidth]{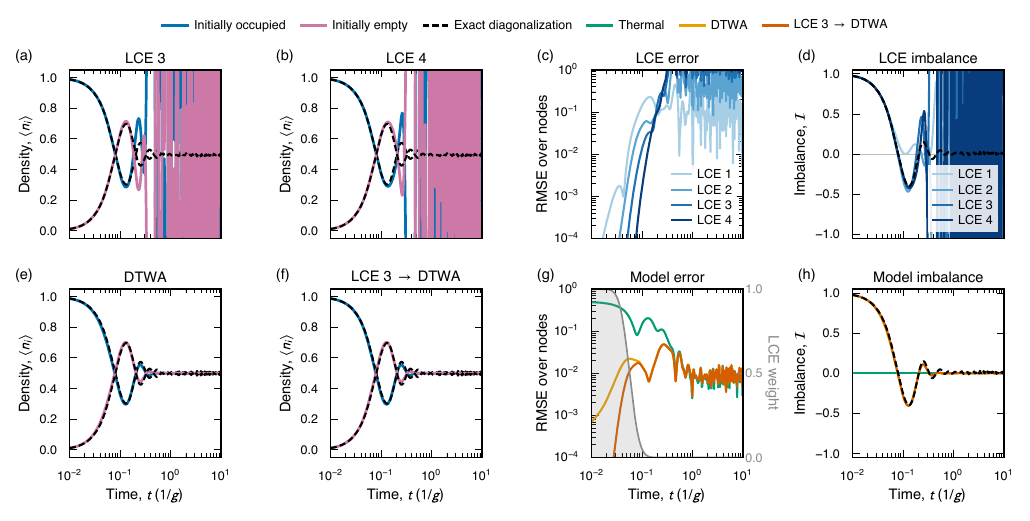}
    \caption{The excitation-cluster expansion and the DTWA fail in opposite regimes, and joining them follows exact dynamics at all times. Checkerboard initial state of a resonant ($\Delta = 0$) $4 \times 4$ transmon array with eight excitations, compared with exact diagonalization (dashed black). (a), (b) Mean occupations of the initially occupied and empty sites from the excitation-cluster expansion of orders 3 and 4 (LCE3, LCE4), which diverge once many-body correlations build up. (c), (d) RMSE over sites relative to exact dynamics and imbalance for LCE orders 1--4. (e), (f) As (a), (b) for the DTWA and for LCE3 joined to the DTWA at $t^*$. (g) RMSE of the DTWA, the thermal occupations and the joined result (left axis), and the LCE weight $w(t)$ (gray, right axis). (h) The imbalance computed with the same methods and of exact dynamics; the DTWA misses the weak late-time fluctuations of the exact curve.}
    \label{fig:lce_dtwa_validation_resonant}
\end{figure*}

We validate the dynamics in \figref{fig:analog_simulation}, the excitation-cluster expansion~\cite{rigolNumericalLinkedClusterApproach2006, tangShortIntroductionNumerical2013, yangQuantumManybodyTheory2008} joined to the DTWA~\cite{polkovnikovPhaseSpaceRepresentation2010, schachenmayerManyBodyQuantumSpin2015} as described in \secref{sec:dynamics}, against exact evolution of a $4 \times 4$ transmon array with homogeneous coupling $g/2\pi = \SI{10}{\mega\hertz}$ and staggered anharmonicities, without couplers, in the full bosonic sector of eight excitations, starting from the checkerboard at resonance (\figref{fig:lce_dtwa_validation_resonant}). The two methods fail in opposite ways, which is why we combine them.

The excitation-cluster expansion is accurate at short times, but fails abruptly as soon as many-body correlations build up. At early times, its error lies orders of magnitude below that of the DTWA and decreases with the order of the expansion [\figpanels{fig:lce_dtwa_validation_resonant}{a}{c}]. The root-mean-square error (RMSE) of the third-order expansion, LCE3, reaches 0.01 at $gt \approx 0.1$, however, and exceeds unity by $gt \approx 0.3$, with unphysical densities and an imbalance of magnitude well above one thereafter [\figpanels{fig:lce_dtwa_validation_resonant}{a}{d}]. Raising the order does not help: LCE4 breaks down at nearly the same time, because the dynamics by then involve correlations among more excitations than any low-order cluster contains.

The DTWA behaves in the opposite way. It is less accurate at short times, with an RMSE about seven times that of LCE3 for $0.03 \leq gt < 0.1$, but it remains stable at all times and reproduces the relaxation, the overshoot of the imbalance, and the thermal plateau, with a largest RMSE of 0.05 [\figpanels{fig:lce_dtwa_validation_resonant}{e}{h}]. Being semiclassical, it misses the weak quantum fluctuations around the plateau: for $gt \geq 1$, the imbalance of the exact dynamics (dashed black) fluctuates about four times more strongly than that of the DTWA (standard deviation 0.008 against 0.002), and the late-time RMSE of 0.009 of the DTWA is no lower than that of the thermal occupations and is dominated by these fluctuations [\figpanel{fig:lce_dtwa_validation_resonant}{g}].

We find heuristically that joining the two methods at the trust time $t^*$ of \secref{sec:dynamics}, here $gt^* = 0.21$, gives the most faithful model overall for the quench dynamics. It inherits the precision of LCE3 at early times and the stability of the DTWA at late times, and its error never substantially exceeds that of the better method: its largest RMSE, reached near the handover, is 0.05, and its largest misplaced fraction is also 0.05 [\figpanels{fig:lce_dtwa_validation_resonant}{f}{h}]. The approach behaves in the same way for a staggered detuning of $\Delta = 55g$, where all time scales are longer (Supplementary Information; \secref{sec:si_dispersive_validation}).

\subsection{Use of large language models and agentic tools}
\label{sec:ai_use}

Large language models (Gemini Pro and Gemini Flash, versions 3.0 and 3.1, Google; Claude Sonnet and Claude Opus, versions 4.5 to 5.5, Anthropic) were used to write the plotting scripts for the data panels of the figures, to extend existing analysis and validation code written by the authors, and to write the scripts for the simulations on high-performance computing resources. They were also used to discuss the choice of methods, to draft sentences or paragraphs of the manuscript and to check analytical derivations. They were not used for the core CUT code or for the schematic illustrations in the figures, which were drawn by the authors. All AI-generated drafts were reviewed and edited by the authors, who take full responsibility for the content of the paper.

\section*{Data availability}

The data underlying the figures are available from the corresponding author upon reasonable request.

\newpage
\bibliography{refs}

\begin{acknowledgments}
    We thank Agustin Di Paolo, Ariadna Soro, and Oskar Watsfeldt for insightful discussions. We acknowledge support from the Knut and Alice Wallenberg Foundation through the Wallenberg Centre for Quantum Technology (WACQT) and from the Horizon Europe programme HORIZON-CL4-2022-QUANTUM-01-SGA via the project 101113946 OpenSuperQPlus100. E.D.\ gratefully acknowledges funding from the Swedish Foundation for Strategic Research via the SwedNESS graduate school (GSn15-0008). A.F.K.\ is also supported by the Swedish Foundation for Strategic Research (grant numbers FFL21-0279 and FUS21-0063) and the Norwegian Research Council through the Norwegian Quantum Software Center (NorQSoft, project number 361350). Computational resources were provided on the Berzelius system funded by the Knut and Alice Wallenberg foundation and operated by the National Academic Infrastructure for Supercomputing in Sweden (NAISS). Computational resources were also provided by NAISS at C3SE, partially funded by the Swedish Research Council through grant agreement number 2022-06725.
\end{acknowledgments}

\section*{Author contributions}
S.P.F.\ conceived the project, developed the theory and the code, and wrote the manuscript. S.P.F.\ and E.D.\ implemented and performed the high-performance computing simulations. J.F.P.\ and A.F.K.\ supervised the project. All authors discussed the results during every stage of the project and reviewed the manuscript.

\clearpage
\onecolumngrid
\begin{center}
    {\large\bfseries Supplementary Information}
\end{center}
\vspace{\baselineskip}
\twocolumngrid

\setcounter{section}{0}
\renewcommand{\theHsection}{SI.\arabic{section}}

\section{Proof of unidirectional flow}
\label{SI:unidirectional}

We prove the statement of \secref{sec:design} that the excitation-conserving flow of the $K$-body couplings depends only on couplings of at most $K$ bodies. We decompose the Hamiltonian and the generator into normal-ordered $K$-body terms, $H = \sum_{K \geq 1} H^{[K]}$ and $\eta = \sum_{K \geq 1} \eta^{[K]}$, with
\begin{equation}
    H^{[K]} = \sum_{IJ} H_{IJ}\, a_{i_1}^\dag \cdots a_{i_K}^\dag a_{j_1} \cdots a_{j_K},
\end{equation}
where $I = (i_1, \dots, i_K)$ and $J = (j_1, \dots, j_K)$ are multi-indices. The tensors $H_{ij}$ and $H_{ijkl}$ of \eqref{eq:device_hamiltonian} are the cases $K = 1, 2$; a constant term commutes with all operators and does not enter the flow. At the end of the proof, we show with a counterexample that the statement does not hold for the non-excitation-conserving flow.

The proof rests on a selection rule for commutators of normal-ordered terms. Normal ordering the product $AB$ of two normal-ordered operators requires moving the annihilation operators of $A$ to the right of the creation operators of $B$. Each such exchange, $a_i a_j^\dag = a_j^\dag a_i + \delta_{ij}$, either keeps both operators or replaces them by a Kronecker delta, i.e., a contraction. Wick's theorem~\cite{kehreinFlowEquationApproach2006} collects the result into normal-ordered terms with $c$ contractions, each of which removes one annihilation operator of $A$ and one creation operator of $B$. For a $p$-body $A^{[p]}$ and a $q$-body $B^{[q]}$, at most $\min(p, q)$ contractions are possible, and a term with $c$ contractions retains $p + q - c$ creation and as many annihilation operators, i.e., it is $(p + q - c)$-body. The term without contractions is the same for $AB$ and $BA$, as bosonic operators commute inside a normal-ordered product, and cancels in the commutator. Hence
\begin{equation}
    \mleft[ A^{[p]}, B^{[q]} \mright] = \sum_{K = \max(p, q)}^{p + q - 1} C^{[K]},
    \label{eq:selection_rule}
\end{equation}
i.e., a commutator never produces a term with fewer bodies than its larger factor.

Since normal-ordered monomials are linearly independent, the flow equation can be projected onto its $K$-body part, and by \eqref{eq:selection_rule},
\begin{equation}
    \frac{\rd H^{[K]}}{\rd l} = \sum_{\substack{p, q \leq K \\ p + q > K}} \mleft[ \eta^{[p]}, H^{[q]} \mright]^{[K]} ,
    \label{eq:hierarchy}
\end{equation}
which contains no Hamiltonian or generator term of more than $K$ bodies. The generator obeys the same hierarchy: $\eta^{[K]}$ depends only on $H^{[K' \leq K]}$. For the sign generator, $\eta^{[K]}_{IJ} = \mathrm{asgn}(E_I - E_J)\, V_{IJ}$, where $E_I$ is the diagonal energy of the $K$-excitation Fock state $a_{i_1}^\dag \cdots a_{i_K}^\dag \ket{0}$. A $K'$-body term contains $K'$ annihilation operators and therefore annihilates every state with fewer than $K'$ excitations, which leaves $E_I$ dependent only on $H^{[K' \leq K]}$. For the Wegner generator, the same follows from \eqref{eq:selection_rule} applied to $\eta = [H_0, V]$. Equation~(\ref{eq:hierarchy}) is therefore a triangular system,
\begin{equation}
    \frac{\rd H^{[K]}}{\rd l} = F_K \mleft( H^{[1]}, \dots, H^{[K]} \mright),
\end{equation}
starting from the closed one-body equation $\rd H^{[1]} / \rd l = [\eta^{[1]}, H^{[1]}]$. Few-body couplings drive many-body couplings, never the reverse, and discarding all terms with $K > K_\mathrm{max}$ leaves the equations for $K \leq K_\mathrm{max}$ unchanged. For $K_\mathrm{max} = 2$, the only discarded term is the once-contracted part of $[\eta^{[2]}, H^{[2]}]$, i.e., the three-body term of \eqref{eq:incremental_error}.

The truncated flow is moreover exact in the few-excitation sectors. Let $P_n$ project onto the sector of $n$ excitations and let $H_{K_\mathrm{max}}(l)$ be the solution truncated at $K_\mathrm{max}$ bodies. Since terms with $K > n$ annihilate this sector, the truncated and exact Hamiltonians and generators coincide on it. Because the flow also conserves excitations, for all $n \leq K_\mathrm{max}$ and at every flow time
\begin{equation}
    P_n H_{K_\mathrm{max}}(l) P_n = P_n H(l) P_n = U(l) P_n H(0) P_n U^\dag(l).
\end{equation}
The truncated flow thus keeps the exact spectrum of all sectors with up to $K_\mathrm{max}$ excitations, and the truncation error first enters at $n = K_\mathrm{max} + 1$. The same counting applies to an operator with $s + 1$ creation and $s$ annihilation operators, such as the flowed creation operator. Its commutator with $\eta^{[p]}$ contains at least $\max(p - 1, s)$ annihilation operators. The flow of its coefficients with $s$ annihilation operators thus involves only coefficients with at most $s$ annihilation operators and generator terms of at most $s + 1$ bodies, and the coefficients with up to $K_\mathrm{max} - 1$ annihilation operators are closed under a generator truncated at $K_\mathrm{max}$ bodies. With $K_\mathrm{max} = 2$, the one- and two-excitation energies and the coefficients $\alpha_{im}$ and $\alpha_{imnl}$ of \secref{sec:locality} are then exact, up to the residual off-diagonal couplings at the end of the flow.

The selection rule, however, relies on each term having as many creation as annihilation operators, which fails for the non-excitation-conserving couplings: the number of contractions is then no longer bounded by half the number of operators in each factor. For a single mode, for instance, $[a^{\dag 3} a, a^\dag a^3] = -6 a^\dag a - 18 a^{\dag 2} a^2 - 8 a^{\dag 3} a^3$, where the commutator of two four-operator terms feeds the one-body coupling. The non-excitation-conserving flow thus feeds back into lower orders, and its truncation is controlled instead by the scale separation $E_{ij} \gg |V^\mathrm{nec}_{ij}|$. The exactness shown above refers to the excitation-conserving Hamiltonian $H^\mathrm{ec}$, not to the device Hamiltonian.

\section{Hard-core bosons and the flowed creation operator}
\label{sec:si_hardcore}

The diagonal model of \eqref{eq:diagonal_hamiltonian} evolves any state given in its eigenbasis trivially. A quench, however, starts from a Fock state of the device, $\ket{\Omega} = \prod_{j \in \mathcal{O}} a_j^\dag \ket{0}$, with $\mathcal{O}$ the set of initially occupied sites, and is read out in the site occupations $n_i(t)$. Unlike the infinite-temperature correlators of Ref.~\cite{thomsonUnravellingQuantumDynamics2024}, which the diagonal route of \secref{sec:analog} yields by sampling product eigenstates without transforming any state, these occupations require mapping the device operators onto the eigenbasis through the flowed creation operator of \secref{sec:locality},
\begin{equation}
    a_i^\dag = \sum_m \alpha_{im} \tilde{a}_m^\dag + \sum_{mnl} \alpha_{imnl} \tilde{a}_m^\dag \tilde{a}_n^\dag \tilde{a}_l + \dots .
    \label{eq:si_flowed_creation}
\end{equation}
Here $\tilde{a}_m^\dag$ creates an excitation in the single-excitation eigenmode $m$, which we label by the site it is continuously connected to during the flow, and $\ket{\tilde{0}}$ is the common vacuum. The linear coefficients $\alpha_{im}$ form the unitary matrix of single-excitation eigenvectors, so $|\alpha_{im}| \leq 1$. The non-linear coefficients $\alpha_{imnl}$ act only on states that already hold an excitation, and the ellipses denote terms with two or more annihilators, which we discard. 

\begin{figure*}
    \centering
    \includegraphics[width=\linewidth]{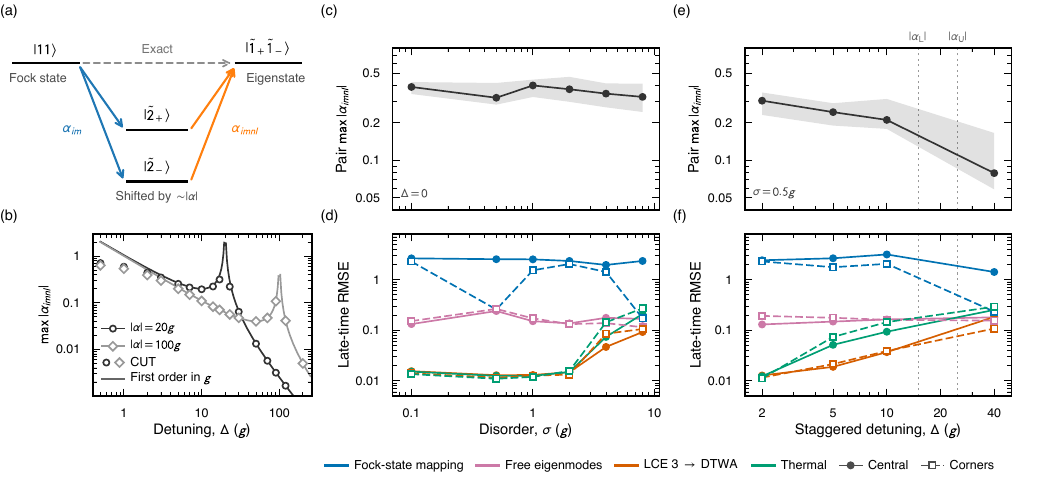}
    \caption{The hard-core constraint limits the Fock-state mapping.
    (a) Two-excitation states of two hybridized transmons. The flow maps the Fock state $\ket{11}$ onto the eigenstate $\ket{\tilde{1}_+ \tilde{1}_-}$ (dashed arrow). The linear coefficients $\alpha_{im}$ alone also place weight on the doubly occupied eigenstates $\ket{\tilde{2}_\pm} = (\tilde{a}_\pm^\dag)^2 \ket{\tilde{0}} / \sqrt{2}$, which the anharmonicity shifts by ${\sim}|\alpha|$, and the non-linear coefficients $\alpha_{imnl}$ must cancel this weight. 
    (b) Largest non-linear coefficient of the dimer as a function of the detuning $\Delta$ for two anharmonicities [markers: CUT; lines: first-order expression, \eqref{eq:si_dimer_coefficient}], peaking at the doublon resonance $\Delta = |\alpha|$. 
    (c), (e) Largest non-linear coefficient with all indices on one nearest-neighbor pair of transmons in a $4 \times 4$ array, as a function of the standard deviation $\sigma$ of the qubit frequencies at $\Delta = 0$ (c) and a staggered detuning $\Delta$ between the sublattices at $\sigma = 0.5g$ (e). Markers show the median and the shading the interquartile range over the 24 pairs and three disorder realizations. Dotted lines mark the anharmonicities $|\alpha_\mathrm{L}| = 15g$ and $|\alpha_\mathrm{U}| = 25g$ of the two sublattices. 
    (d), (f) Root-mean-square error of the site occupations relative to exact dynamics of the same array with four excitations, averaged over all sites and $1.5 \leq gt < 10$, for one realization, starting from the central plaquette (filled markers, solid lines) or the corners (open markers, dashed lines). Shown are the Fock-state mapping, evaluated as described in the text with the flowed creation operator, the interactions of the diagonal model and the inverse transformation truncated at three operators, which does not conserve the excitation number; free eigenmodes, i.e., the same without the non-linear coefficients and the interactions; LCE3$\,\to\,$DTWA; and the high-temperature thermal occupations.}
    \label{fig:ed_renormalizability}
\end{figure*}

Both sets of coefficients are exact up to the residual off-diagonal couplings (Supplementary Information; \secref{SI:unidirectional}), and the discarded terms do not act on states with fewer than two excitations, so \eqref{eq:si_flowed_creation} maps all states with up to two excitations exactly; the truncation enters from three excitations onwards. We call the flow renormalizable for this operator if the expansion is controlled, i.e., if the flow transfers only little weight to operators of higher order, such that the coefficients decrease with the number of operators and the truncation remains accurate for any number of excitations. At the order kept here, a necessary condition is $\max |\alpha_{imnl}| \ll 1$, the quantity shown in \figpanels{fig:ed_renormalizability}{b}{c} and \figpanelNoPrefix{fig:ed_renormalizability}{e}. Below, we first show why this condition fails for transmons and then describe how we evaluate the resulting Fock-state mapping and compare it with exact dynamics.

For transmons with anharmonicity $|\alpha| \gg g$, this condition fails at weak disorder, for a reason that is absent for fermions. Consider two transmons with exchange coupling $g$, detuning $\Delta = \omega_2 - \omega_1 \geq 0$ and equal anharmonicity $\alpha$ [\figpanel{fig:ed_renormalizability}{a}]. Their single-excitation eigenmodes, with energies $\bar{\omega} \pm \frac{1}{2} \sqrt{\Delta^2 + 4g^2}$, are created by $\tilde{a}_\pm^\dag$ and hybridize the transmons with the mixing angle $\theta$ ($\tan 2\theta = 2g/\Delta$), such that the linear part of \eqref{eq:si_flowed_creation} reads $a_1^\dag = \cos\theta\, \tilde{a}_-^\dag + \sin\theta\, \tilde{a}_+^\dag$ and $a_2^\dag = -\sin\theta\, \tilde{a}_-^\dag + \cos\theta\, \tilde{a}_+^\dag$. Among the two-excitation eigenstates, $\ket{\tilde{1}_+ \tilde{1}_-} = \tilde{a}_+^\dag \tilde{a}_-^\dag \ket{\tilde{0}}$ holds one excitation in each eigenmode, whereas the doubly occupied states $\ket{\tilde{2}_\pm} = (\tilde{a}_\pm^\dag)^2 \ket{\tilde{0}} / \sqrt{2}$ hold both in the same eigenmode and are shifted by ${\sim}|\alpha|$. In the hard-core regime, $g \ll |\alpha|$ and $\Delta < |\alpha|$, the flow maps the device state $\ket{11} = a_1^\dag a_2^\dag \ket{0}$ onto $\ket{\tilde{1}_+ \tilde{1}_-}$, which keeps the two excitations on different transmons. The linear coefficients alone instead yield
\begin{equation}
    \cos 2\theta \ket{\tilde{1}_+ \tilde{1}_-} + \frac{\sin 2\theta}{\sqrt{2}} \mleft( \ket{\tilde{2}_+} - \ket{\tilde{2}_-} \mright),
\end{equation}
which places the weight $\sin^2 2\theta = 4g^2 / (\Delta^2 + 4g^2)$ on the doubly occupied states, all of it at resonance. 

The non-linear coefficients must cancel this weight in the hard-core regime. Comparing with the flow of the dimer, we find, to first order in $g$,
\begin{equation}
    \max |\alpha_{imnl}| \approx \left| \frac{g}{\Delta} + \frac{g}{|\alpha| - \Delta} \right| = \frac{g |\alpha|}{\Delta\, \big| |\alpha| - \Delta \big|} ,
    \label{eq:si_dimer_coefficient}
\end{equation}
within \SI{6}{\percent} for $\Delta \geq 5g$ at $|\alpha| = 20g$ and $100g$, except within ${\sim}3g$ of the resonance at $|\alpha| = 20g$, where it overestimates by about \SI{20}{\percent} [\figpanel{fig:ed_renormalizability}{b}]. For $\Delta \lesssim g$, where \eqref{eq:si_dimer_coefficient} diverges, the coefficient instead saturates at 0.5--0.7, as large as the linear coefficients. The first term of \eqref{eq:si_dimer_coefficient} cancels the weight that the mixing places on $\ket{\tilde{2}_\pm}$, and the second is the genuine admixture of doubly occupied states, which diverges at the doublon resonance $\Delta = |\alpha|$. In the hard-core regime, $g \ll \Delta \ll |\alpha|$, the non-linear coefficients are thus of first order in the mixing angle, $\theta \approx g/\Delta$, as large as the single-particle dressing itself; only for $\Delta \gg |\alpha|$, where the bosons become effectively free, do they fall off as $g|\alpha|/\Delta^2$. For fermions, exclusion is built into the operator algebra; for transmons, it is imposed dynamically by the anharmonicity and, expressed in a free-mode basis, must be carried by non-linear terms at every order in the number of excitations.

In an array, every pair of coupled transmons forms such a dimer. For each nearest-neighbor pair $(i, j)$ of a $4 \times 4$ array of directly coupled transmons (anharmonicities $15g$ and $25g$ on the two sublattices), we therefore take the largest $|\alpha_{kmnl}|$ with all four indices in $\{i, j\}$, the dimer coefficient of \figpanel{fig:ed_renormalizability}{b} embedded in the array. Its median stays between $0.3$ and $0.4$ for disorder strengths up to $\sigma = 8g$ [\figpanel{fig:ed_renormalizability}{c}] and falls below $0.1$ only for a staggered detuning beyond the anharmonicities of both sublattices [\figpanel{fig:ed_renormalizability}{e}]. Near-resonant pairs, $|\Delta_{ij}| \lesssim g$ or $|\Delta_{ij}| \approx |\alpha|$, dominate the largest coefficients at any disorder strength. The flow is therefore renormalizable only once the detuning exceeds the anharmonicity.

We evaluate the Fock-state mapping in the Heisenberg picture, which is equivalent to mapping $\ket{\Omega}$ onto the eigenbasis by applying \eqref{eq:si_flowed_creation} once per excitation. The occupations $n_i(t) = \bra{\Omega} a_i^\dag(t)\, a_i(t) \ket{\Omega}$ require the device creation operator at time $t$, which we obtain in three steps, each truncated at three operators, consistently with \eqref{eq:si_flowed_creation}. First, we express $a_i^\dag$ in the eigenbasis through \eqref{eq:si_flowed_creation}. Second, we evolve the eigenmode operators with the diagonal Hamiltonian. Let $E_m$ be the energy of a single excitation in eigenmode $m$, and $\zeta_{mn}$ the interaction energy of two excitations in eigenmodes $m$ and $n$, with $\zeta_{mm}$ that of a doubly occupied eigenmode. Then $\tilde{a}_m^\dag(t) = \e^{\mathrm{i} E_m t}\, \tilde{a}_m^\dag \exp(\mathrm{i} t \sum_n \zeta_{mn} \tilde{n}_n)$, and expanding the exponential to first order in the number operators gives
\begin{equation}
    \tilde{a}_m^\dag(t) \approx \e^{\mathrm{i} E_m t} \Big[ \tilde{a}_m^\dag + \sum_n \mleft( \e^{\mathrm{i} \zeta_{mn} t} - 1 \mright) \tilde{a}_m^\dag \tilde{n}_n \Big],
    \label{eq:si_diagonal_evolution}
\end{equation}
whereas the three-operator terms of \eqref{eq:si_flowed_creation} acquire only the phases $\e^{\mathrm{i} (E_m + E_n - E_l) t}$. Third, we return to the device basis with the inverse of \eqref{eq:si_flowed_creation} to first order in the non-linear coefficients,
\begin{equation}
    \tilde{a}_m^\dag = \sum_j \beta_{mj}\, a_j^\dag + \sum_{jkl} \beta_{mjkl}\, a_j^\dag a_k^\dag a_l + \dots ,
    \label{eq:si_inverse}
\end{equation}
with $\beta = \alpha^{-1}$ and
\begin{equation}
    \beta_{mjkl} = - \sum_{inpq} \beta_{mi}\, \alpha_{inpq}\, \beta_{nj}\, \beta_{pk}\, \beta_{ql}^* .
\end{equation} 

The result has the form $a_i^\dag(t) = \sum_j A_{ij}(t)\, a_j^\dag + \sum_{jkl} A_{ijkl}(t)\, a_j^\dag a_k^\dag a_l$, with the linear part $A_{ij}(t) = \sum_m \alpha_{im} \e^{\mathrm{i} E_m t} \beta_{mj}$, and we evaluate its occupations in the Fock state $\ket{\Omega}$. The truncated operator no longer obeys the bosonic commutation relations, so the total occupation $\sum_i n_i(t)$ is not conserved. As a reference that conserves it, we also evaluate free eigenmodes, i.e., the same construction without the non-linear coefficients and the interactions ($\alpha_{imnl} = \zeta_{mn} = 0$). Then $A_{ijkl} = 0$ and
\begin{equation}
    n_i(t) = \sum_j |A_{ij}(t)|^2\, n_j(0) ,
\end{equation}
which describes non-interacting bosons with the single-excitation spectrum of the array and conserves the excitation number, since $\alpha_{im}$ is unitary.

Finally, we compare the Fock-state mapping and the free eigenmodes with exact dynamics of the same $4 \times 4$ array with four excitations [\figpanel{fig:ed_renormalizability}{d} and \figpanelNoPrefix{fig:ed_renormalizability}{f}]. The Fock-state mapping violates excitation-number conservation strongly: the total occupation grows from four to as much as 56, and the late-time error is of order unity for all $\sigma \leq 8g$ and $\Delta \leq 10g$ from the central plaquette and for most of them from the corners. The free eigenmodes neglect the interactions, and their error stays at 0.1--0.3 throughout. The late-time error of LCE3$\,\to\,$DTWA rises from about $0.01$ for $\sigma \leq 2g$ to about $0.1$ at $\sigma = 8g$, where it still lies below those of the thermal occupations, the free eigenmodes and the Fock-state mapping.

\section{Error estimate}
\label{sec:si_error}

We estimate the time up to which the diagonal route of \secref{sec:analog} can be trusted by adapting the incremental truncation-error bound of Ref.~\cite{thomsonUnravellingQuantumDynamics2024} to the sign generator and to transmons. We use the notation of \sisecref{SI:unidirectional}, in which $X^{[K]}$ denotes the normal-ordered $K$-body part of an operator $X$. At each flow step, discarding the three-body terms, i.e., the once-contracted part of the commutator of the two-body parts of the generator and the Hamiltonian, incurs the error
\begin{equation}
    \| A(l) \|_2 = \mleft\| \mleft[ \eta^{[2]}(l), H^{[2]}(l) \mright]^{[3]} \mright\|_2 \lesssim \sqrt{2}\, \| \eta^{[2]}(l) \|_2 \, \| H^{[2]}(l) \|_2 ,
    \label{eq:incremental_error}
\end{equation}
where, for the sign generator, $\| \eta^{[2]} \|_2 = \| V^{[2]} \|_2$, with $V^{[2]}$ the off-diagonal two-body couplings. For a transmon array with exchange coupling $g$, anharmonicity $\alpha$ and disorder strength $d$, the two-body part of the Hamiltonian is dominated by the anharmonicity, $\| H^{[2]} \| \sim |\alpha|$ per site, while the off-diagonal one-body couplings decay as $V^{[1]}(l) \sim g\, \e^{-d l}$, since the sign generator damps each coupling at a rate set by its energy difference (cf.~$\rd V_{ij}/\rd l = -|E_i - E_j| V_{ij}$). Off-diagonal two-body couplings are not present initially but are generated by the flow, predominantly as doublon-forming terms sourced by $[\eta^{[1]}, H_0^{[2]}] \sim |\alpha|\, V^{[1]}(l)$, and are damped at the rate $\Lambda \sim \max(d, |\alpha|)$ set by the two-excitation energy differences they connect. Hence,
\begin{equation}
    \| V^{[2]}(l) \| \sim g |\alpha| \, \frac{\e^{-d l} - \e^{-\Lambda l}}{\Lambda - d},
\end{equation}
and integrating \eqref{eq:incremental_error} over the flow gives the total truncation error
\begin{equation}
    \varepsilon_T \lesssim \sqrt{2}\, |\alpha| \int_0^\infty \rd l \, \| V^{[2]}(l) \| \sim \frac{g \alpha^2}{d \, \Lambda} .
\end{equation}
Energy differences below $\varepsilon_T$ are not resolved, so the diagonal route is limited to times $T \sim 1/\varepsilon_T$. For weak anharmonicity, $d \gg |\alpha|$, this gives $T \sim d^2/(g \alpha^2)$, which recovers the result of Ref.~\cite{thomsonUnravellingQuantumDynamics2024}, $T \sim d^2/(J \Delta_0^2)$, with the hopping $J \to g$ and the interaction $\Delta_0 \to \alpha$. In the hard-core regime of transmon arrays, $g \ll d \ll |\alpha|$, the doublon-forming terms relax quickly but are sourced throughout the flow, and
\begin{equation}
    T \sim \frac{d}{g |\alpha|}, \qquad gT \sim \frac{d}{|\alpha|} \ll 1,
\end{equation}
i.e., the truncated diagonal route cannot be trusted beyond a single hopping time unless the disorder exceeds the anharmonicity. The estimate mirrors the magnitude of the non-linear coefficients of the flowed creation operator, $|\alpha_{imnl}| \approx |g/\Delta + g/(|\alpha| - \Delta)|$ [\figref{fig:ed_renormalizability}(b)], which is of first order in the mixing angle for $\Delta \ll |\alpha|$ and of order $g|\alpha|/\Delta^2$ only for $\Delta \gg |\alpha|$. As in Ref.~\cite{thomsonUnravellingQuantumDynamics2024}, the estimate concerns typical rather than worst-case couplings and neglects near-degenerate pairs, which only shorten $T$; unlike for the fermionic models of that work, the comparison with exact dynamics [\figpanel{fig:ed_renormalizability}{d} and \figpanelNoPrefix{fig:ed_renormalizability}{f}] confirms that the breakdown is genuine for transmon arrays.

\section{Transport versus detuning}
\label{sec:si_transport}

\begin{figure}
    \centering
    \includegraphics[width=\linewidth]{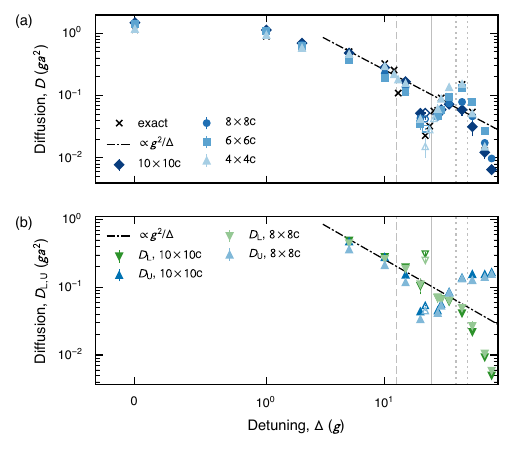}
    \caption{Device-level physics sets the transport between the sublattices. 
    (a) Diffusion constant $D$ of the central cluster versus the staggered detuning $\Delta$ for arrays with couplers of $4 \times 4$ to $10 \times 10$ qubits (LCE3$\,\to\,$DTWA, as in \figref{fig:analog_simulation}), with the second-order law $D \propto g^2/\Delta$ (dash-dotted) and the same fit to exact dynamics of the effective model of the $4 \times 4$ array (crosses). For $\Delta \geq 5g$, $D$ is fitted to diffusion on each sublattice; for $\Delta \leq 2g$, it is the diffusion constant of a telegrapher fit (ballistic to diffusive crossover). 
    (b) Diffusion constants $D_\mathrm{L}$ and $D_\mathrm{U}$ of the excitations on the lower (green) and upper (blue) sublattice, fitted separately, for the $8 \times 8$ (light) and $10 \times 10$ (dark) arrays: above ${\sim}40g$, $D_\mathrm{U}$ saturates and $D_\mathrm{L}$ collapses [the $10 \times 10$ data are shown in \figpanel{fig:analog_simulation}{g}]. Vertical lines: doublon resonances of the anharmonicities $-15g$ and $-25g$, at first order ($\Delta = 25g$, solid), second order ($12.5g$, dashed) and higher orders ($40g$, $50g$, dotted). Open markers: second-order hard-core projection of the DTWA unreliable (largest ratio of doublon-formation amplitude to energy denominator above $0.3$).}
    \label{fig:ed_detuning_sweep}
\end{figure}

\figref{fig:ed_detuning_sweep} shows the diffusion constants behind \figpanel{fig:analog_simulation}{g}, fitted as described in Methods, \secref{sec:transport_fits}. Since the second-order hopping of the staggered array has no bonds between the sublattices, each sublattice conserves its excitations, and $D_\mathrm{L}$ and $D_\mathrm{U}$ are fitted separately for $\Delta \geq 5g$. The combined fit of both sublattices is not a converged quantity at large detuning: at $\Delta = 55g$, it gives $D = 0.079$, 0.046, and $0.032\,ga^2$ for arrays of $6 \times 6$, $8 \times 8$, and $10 \times 10$ qubits, respectively, because its window moves towards the slower sublattice as the array grows. The sublattice-resolved values, in contrast, agree between the two largest arrays to within about \SI{20}{\percent}. Excitations hop via virtual doubly occupied states: at $\Delta = 55g$, neglecting the doublon-forming terms of the effective model overestimates $D$ by ${\sim}\SI{25}{\percent}$ for a $4 \times 4$ array with couplers.

\begin{figure*}
    \centering
    \includegraphics[width=\linewidth]{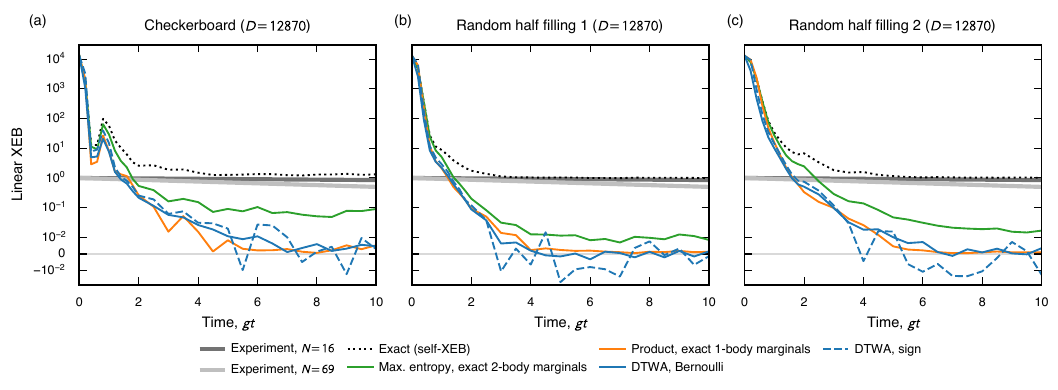}
    \caption{Bitstrings sampled from the DTWA lose their cross-entropy once the dynamics scramble. Linear cross-entropy benchmark (XEB) against the exact bitstring distribution of a $4 \times 4$ array with couplers at resonance and half filling (dimension $D = 12\,870$), for (a) the checkerboard and (b), (c) two random initial Fock states: DTWA trajectories read by the sign of $z_i$ (dashed blue) or by Bernoulli draws (solid blue); post-selected product of the exact densities (orange); maximum-entropy distribution of the exact one- and two-body marginals (green). Dotted: exact self-XEB (Porter--Thomas value 1). Gray lines: fidelity $\mathrm{e}^{-\epsilon N g t}$ for the error rate $\epsilon = \SI{0.1}{\percent}$ per qubit per $1/g$ of Ref.~\cite{andersenThermalizationCriticalityAnalogue2025a}, for $N = 16$ (dark) and $69$ (light) qubits. The vertical axis is linear for $|F| < 0.01$ and logarithmic above.}
    \label{fig:si_xeb}
\end{figure*}

\section{Cross-entropy of bitstrings sampled from the DTWA}
\label{sec:si_xeb}

Cross-entropy benchmarking (XEB) compares measured bitstrings $x$ with the ideal distribution $p(x, t) = |\langle x | \psi(t) \rangle|^2$ through the linear XEB $F = D \sum_x q(x)\, p(x) - 1$, where $q$ is the distribution of the sampled bitstrings and $D$ the dimension of the excitation sector; $F = 0$ for uniform sampling within the sector and $F \approx 1$ for ideal sampling of a Porter--Thomas distribution~\cite{andersenThermalizationCriticalityAnalogue2025a}. To test whether our approach could produce such samples, we compare four classical samplers with the exact $p(x, t)$ of the $4 \times 4$ array with couplers at resonance, at half filling ($n^\mathrm{tot} = 8$, $D = 12\,870$), for the checkerboard and two random initial Fock states (\figref{fig:si_xeb}). Exact dynamics and samplers use the same hard-core model as the DTWA in the main text, including the doublon-mediated interaction and correlated hopping, such that the comparison isolates the sampling. All samplers are post-selected on the number of excitations, as in the experiment. From each of $2 \times 10^4$ DTWA trajectories, we draw one bitstring either by reading each qubit as excited if $z_i > 0$ or by drawing it with probability $(1 + z_i)/2$; for the latter, $q$ is evaluated exactly as the average over trajectories of the post-selected product distributions. As references for samplers that reproduce low-order marginals exactly, we use the post-selected product of the exact densities $n_i$ and the maximum-entropy distribution on the sector that reproduces the exact $\langle n_i \rangle$ and $\langle n_i n_j \rangle$.

Before the bitstring distribution has scrambled, $gt \lesssim 2$, the low-order marginals carry most of the cross-entropy, and all samplers reach large values of $F$. Once the self-XEB of the exact distribution, $D \sum_x p^2 - 1$, approaches its Porter--Thomas value of one, at $gt \approx 3$--$4$, the cross-entropy of the DTWA samplers falls to a few hundredths and, for $gt \geq 5$, to about $0.01$ (for the sign readout, within its sampling uncertainty of $\pm 0.02$), comparable to the product of exact densities and below the maximum-entropy distribution of exact two-body marginals ($0.01$--$0.1$). By contrast, the error rate of \SI{0.1}{\percent} per qubit per $1/g$ reported in Ref.~\cite{andersenThermalizationCriticalityAnalogue2025a} implies fidelities $\mathrm{e}^{-0.001 N g t}$ of 0.5--0.95 over this time window for $N = 16$ to 69 qubits. Sampling from the DTWA, or from the least-biased distribution consistent with the exact two-body marginals, thus falls short of the experiment by about two orders of magnitude in the regime of its beyond-classical claims, and we expect the gap to widen with system size, as the weight of low-order marginals in the cross-entropy decreases with $D$. The maximum-entropy distribution measures the cross-entropy carried by the two-body marginals alone; it is not an upper bound, since distributions with the same marginals but additional structure, ultimately $p$ itself, can reach higher values. The test therefore concerns our sampler and does not bear on other classical algorithms.

\section{DTWA with and without the couplers}
\label{sec:si_fullarray_dtwa}
\begin{figure*}
    \centering
    \includegraphics[width=\linewidth]{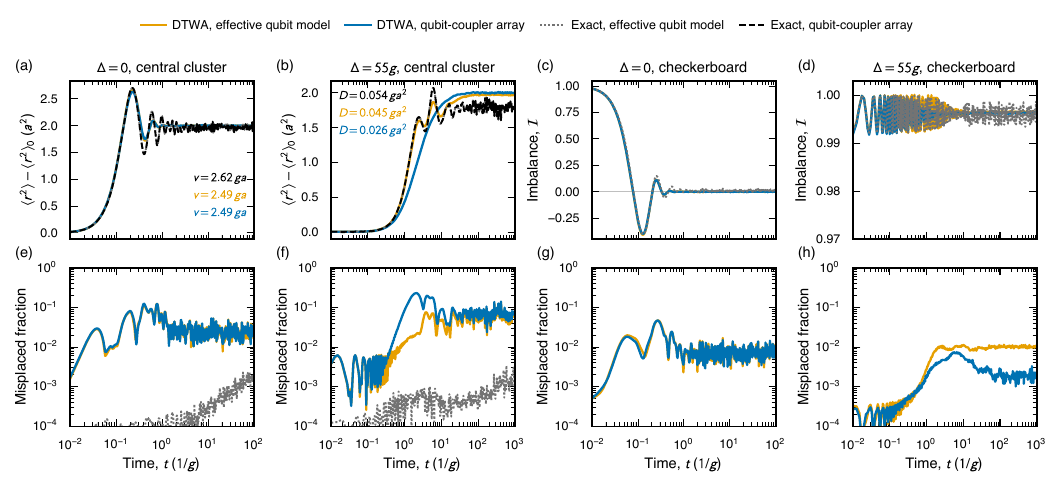}
    \caption{Eliminating the couplers before sampling makes the DTWA cheaper and, at large detuning, more accurate. Calibrated $4 \times 4$ array with couplers of \figref{fig:analog_simulation} at resonance and for $\Delta = 55g$: DTWA of the effective qubit model (orange, 16 modes) and of the qubit-coupler array in its dressed basis (blue, 40 modes), with exact dynamics of the array (dashed black) and of the effective model (dotted gray). 
    (a), (b) Mean-square displacement of a central cluster of four excitations, with the fitted velocity $v$ and diffusion constant $D$ of each method (Methods; \secref{sec:transport_fits}). 
    (c), (d) Imbalance of the checkerboard with eight excitations. 
    (e)--(h) Misplaced fraction $\sum_i |n_i - n_i^\mathrm{ref}| / 2n^\mathrm{tot}$ of the excitations relative to a reference $n_i^\mathrm{ref}$: exact dynamics of the array for the central cluster (e, f), whose dotted gray line is the error of the effective model itself, and exact dynamics of the effective model for the checkerboard (g, h), whose sector of the array is out of reach.}
    \label{fig:si_fullarray_dtwa}
\end{figure*}

The DTWA can also be applied to the device model directly, treating each coupler as a further hard-core mode. We compare this with the DTWA of the effective qubit model for the calibrated $4 \times 4$ array with couplers of \figref{fig:analog_simulation} (\figref{fig:si_fullarray_dtwa}). To compare excitation-conserving models, we take the array after the first flow, which eliminates only the non-excitation-conserving terms, and rotate it to the basis that block-diagonalizes its one-body part between qubits and couplers, so that initial states and observables are dressed qubit modes, as in the effective model. Both DTWAs use the same hard-core projection, $2 \times 10^4$ trajectories, and integration steps that resolve the fastest precession. For the central cluster, we compute exact dynamics of all 40 modes in the sector of four excitations (123\,410 states); for the checkerboard, whose sector is out of reach, the reference is the exact dynamics of the effective model.

The effective model is accurate: its exact dynamics reproduce those of the array with a largest misplaced fraction of 0.002 ($\Delta = 0$) and 0.003 ($\Delta = 55g$) [\figpanels{fig:si_fullarray_dtwa}{e}{f}], so the error of the DTWA of the effective model is that of the DTWA alone. At resonance, the two DTWAs agree to within their sampling noise (a misplaced fraction of 0.014 between them, against 0.011 between two sets of trajectories), with $v = 2.49\,ga$ against $2.62\,ga$ in exact dynamics. At $\Delta = 55g$, the DTWA of the array delays the spreading of the cluster: its misplaced fraction peaks at 0.23 near $gt \approx 2$, against 0.13 for the effective model, its logarithmic-time mean doubles from 0.032 to 0.061, and its diffusion constant, $(0.026 \pm 0.002)\,ga^2$, is half the exact value of $(0.054 \pm 0.008)\,ga^2$, whereas the effective model gives $(0.045 \pm 0.005)\,ga^2$ [$(0.051 \pm 0.003)\,ga^2$ for LCE3$\,\to\,$DTWA]. For the checkerboard, both DTWAs follow the exact dynamics of the effective model to a largest misplaced fraction of 0.05 ($\Delta = 0$) and 0.01 ($\Delta = 55g$) and reproduce the frozen imbalance at $\Delta = 55g$. Treating the couplers explicitly thus gains no accuracy, and loses it where transport rests on weak second-order couplings. It also costs more: the couplers, about \SI{1}{\giga\hertz} above the qubits, require 4--14 times more integration steps, and the DTWA of the 40 modes took 13 ($\Delta = 55g$) to 43 ($\Delta = 0$) times longer on a GPU than that of the 16-mode effective model.

\section{Validation of the dynamics at large detuning}
\label{sec:si_dispersive_validation}

\begin{figure*}
    \centering
    \includegraphics[width=\linewidth]{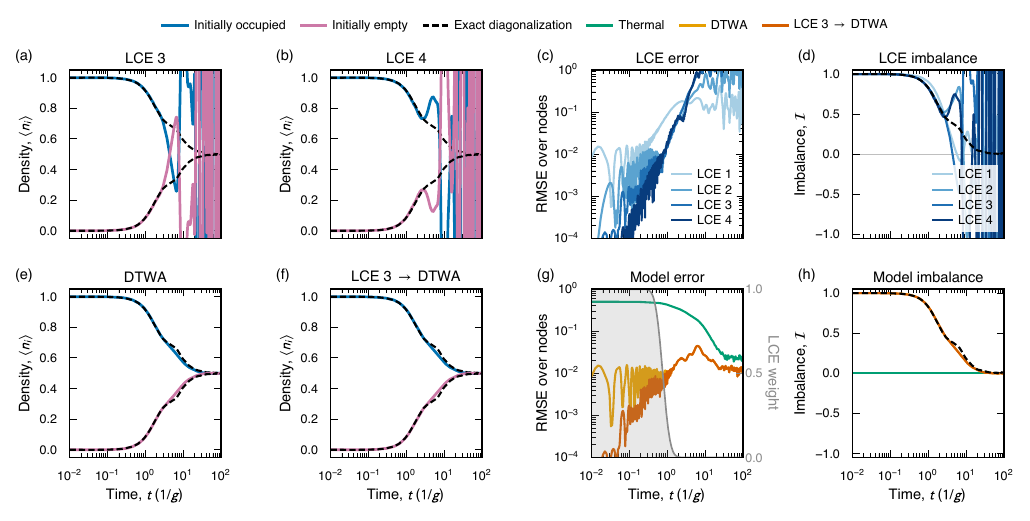}
    \caption{At a staggered detuning $\Delta = 55g$, the same complementary failure occurs on longer time scales. Domain-wall initial state of a $4 \times 4$ transmon array with eight excitations; panels as in \figref{fig:lce_dtwa_validation_resonant}.}
    \label{fig:lce_dtwa_validation_dispersive}
\end{figure*}

\figref{fig:lce_dtwa_validation_dispersive} repeats the validation of \secref{sec:lce3_dtwa_validation} for a staggered detuning $\Delta = 55g$. Since a checkerboard on the lower sublattice stays frozen, we start from a domain wall, with the two upper rows of the $4 \times 4$ array excited, which half fills each sublattice and relaxes through second-order hopping. The same picture holds on longer time scales. The RMSE of LCE3 reaches 0.01 at $gt \approx 0.8$ and exceeds unity at $gt \approx 9$, and LCE4 again fails at nearly the same time. The DTWA has an RMSE of about 0.01 at early times, where LCE3 lies up to two orders of magnitude lower, reaches at most 0.045 near $gt \approx 6$, and again underestimates the late-time fluctuations: for $30 \leq gt < 100$, the standard deviation of the imbalance is 0.006 in the exact dynamics and 0.003 in the DTWA. Joined at $gt^* = 2.4$, the combined result has a largest misplaced fraction of 0.035. Around the handover, for $1 \leq gt < 3$, its mean RMSE of 0.03 exceeds that of the DTWA alone (0.02), as LCE3 has already begun to deviate; the trust time is chosen to protect against the breakdown of the expansion, and this error remains small.

\section{Effective temperature of the initial states}
\label{sec:si_temperature}

A quench conserves the number of excitations $n^\mathrm{tot}$ and the energy. If the dynamics thermalize, local observables therefore relax to their values in the canonical ensemble $\rho_\beta \propto P_{n^\mathrm{tot}} \e^{-\beta H} P_{n^\mathrm{tot}}$, with $P_{n^\mathrm{tot}}$ the projector onto the $n^\mathrm{tot}$-excitation sector, at the inverse temperature $\beta$ fixed by the energy of the initial state (we set $\hbar = k_\mathrm{B} = 1$). We compute $\beta$ for the initial states of \figref{fig:analog_simulation} from the hard-core projection of the effective qubit model on which the DTWA runs (Methods; \secref{sec:dynamics}),
\begin{equation}
    H = \sum_i \omega_i n_i + \sum_{i \neq j} g_{ij}\, \sigma_i^+ \sigma_j^- + \sum_{i<j} V_{ij}\, n_i n_j ,
    \label{eq:hard_core_model}
\end{equation}
where $V_{ij}$ contains the ZZ couplings and the interaction mediated by virtual doublons (about $0.2g$ between neighbors at resonance). Correlated hopping has no diagonal matrix elements in the Fock basis and changes the energy variance only at order $g^4/\alpha^2$; we neglect it here. A Fock state has no hopping energy, so its energy is $E_0 = \sum_i \omega_i n_i + \sum_{i<j} V_{ij} n_i n_j$, and $\beta$ solves
\begin{equation}
    \frac{\mathrm{Tr}\big[P_{n^\mathrm{tot}} H \e^{-\beta H}\big]}{\mathrm{Tr}\big[P_{n^\mathrm{tot}} \e^{-\beta H}\big]} = E_0 .
    \label{eq:temperature_condition}
\end{equation}

For initial states in the bulk of the spectrum, \eqref{eq:temperature_condition} can be solved by expanding in $\beta$. To first order, $\langle H \rangle_\beta \approx \langle H \rangle_0 - \beta \sigma_0^2$, where $\langle H \rangle_0$ and $\sigma_0^2$ are the mean and variance of $H$ in the infinite-temperature ensemble of the sector. Both follow exactly from sums over the $\binom{N}{n^\mathrm{tot}}$ Fock configurations, which reduce to closed-form combinatorial factors, and give
\begin{equation}
    \beta \approx -\frac{E_0 - \langle H \rangle_0}{\sigma_0^2} = -\frac{z}{\sigma_0}, \qquad z = \frac{E_0 - \langle H \rangle_0}{\sigma_0} .
\end{equation}
The expansion holds for $|z| \lesssim 1$, i.e., for initial states whose energy lies within the bulk of the spectrum. For the $4 \times 4$ array, we check it against the exact solution of \eqref{eq:temperature_condition} obtained by fully diagonalizing $H$ in the sector (12\,870 states for the checkerboard).

For a staggered detuning $\Delta = 55g$, the checkerboard fills the lower-frequency sublattice L and lies far below the bulk of its sector ($z = -9.9$ for the $10 \times 10$ array), where the expansion fails. It is then close to the dressed ground state of the sector, from which it differs only by second-order hopping,
\begin{equation}
    E_0 - E_\mathrm{gs} \approx \sum_{i \in \mathrm{L},\, j \in \mathrm{U}} \frac{|g_{ij}|^2}{E_{ij} - E_0} \approx 3.3\, g ,
\end{equation}
where $E_{ij}$ is the energy of the Fock state with the excitation on $i$ moved to $j$. The lowest excitations are $m$ particle--hole pairs across the gap $\Delta$, with $\binom{N_\mathrm{L}}{m}\binom{N_\mathrm{U}}{m}$ configurations for $N_\mathrm{L}$ and $N_\mathrm{U}$ sites on the two sublattices, and \eqref{eq:temperature_condition} becomes $\Delta \langle m \rangle_\beta = E_0 - E_\mathrm{gs}$. For $\langle m \rangle_\beta \ll 1$,
\begin{equation}
    \beta \Delta \approx \ln \frac{N_\mathrm{L} N_\mathrm{U}\, \Delta}{E_0 - E_\mathrm{gs}} .
\end{equation}
For the $4 \times 4$ array, this gives $\beta g = 0.163$, which equals the exact solution of \eqref{eq:temperature_condition}.

\begin{table}
    \centering
    \caption{Effective inverse temperature of the initial states of \figref{fig:analog_simulation}. $n^\mathrm{tot}$: number of excitations; $z$: energy of the initial state relative to the infinite-temperature ensemble of its sector, in units of its standard deviation; $\beta g$: from the high-temperature expansion, or from particle--hole excitations across the gap ($^\dagger$); exact: full diagonalization in the sector.}
    \label{tab:temperature}
    \begin{tabular}{lllrrrr}
        \hline\hline
        $\Delta$ & State & Array & $n^\mathrm{tot}$ & $z$ & $\beta g$ & Exact \\
        \hline
        $0$ & Checkerboard & $4 \times 4$c & 8 & $-0.29$ & $0.081$ & $0.081$ \\
        & & $10 \times 10$c & 50 & $-0.87$ & $0.091$ & \\
        & Cluster & $4 \times 4$c & 4 & $0.17$ & $-0.055$ & $-0.055$ \\
        & & $10 \times 10$c & 12 & $0.42$ & $-0.067$ & \\
        $55g$ & Checkerboard & $4 \times 4$c & 8 & $-3.9$ & $0.163^\dagger$ & $0.163$ \\
        & & $10 \times 10$c & 50 & $-9.9$ & $0.193^\dagger$ & \\
        & Cluster & $4 \times 4$c & 4 & ${\approx}\,0$ & ${\approx}\,10^{-4}$ & \\
        & & $10 \times 10$c & 12 & ${\approx}\,0$ & ${\approx}\,10^{-4}$ & \\
        \hline\hline
    \end{tabular}
\end{table}

Table~\ref{tab:temperature} lists the resulting temperatures. At resonance, both initial states lie within one standard deviation of the infinite-temperature energy, $|\beta| g < 0.1$. The checkerboard, with no two excitations on neighboring sites, avoids the doublon-mediated repulsion and has a positive temperature, $T \approx 11g$, whereas the compact cluster pays it and has a negative temperature, $T \approx -15g$. Since the two sublattices are equivalent at resonance, the thermal imbalance vanishes for any $\beta$; the finite temperature only enters correlators at order $\beta g \sim 0.1$, such as nearest-neighbor density correlations. For $\Delta = 55g$, the cluster, with equally many excitations on both sublattices, is at infinite temperature to within $10^{-4}$. The checkerboard, in contrast, is a low-temperature state, $T \approx \Delta/10$ (for the $10 \times 10$ array, $T = 5.2g$): for the $4 \times 4$ array, the dressed ground state carries \SI{99.2}{\percent} of the thermal weight. Its thermal imbalance, $1 - 2\langle m \rangle_\beta/n^\mathrm{tot}$ reduced by the virtual admixture of the dressed eigenstates, $2 (E_0 - E_\mathrm{gs})/(n^\mathrm{tot} \Delta)$, is 0.995 for the $10 \times 10$ array, against 0.996 at late times in the simulation; for the $4 \times 4$ array, the exact canonical value 0.9961 agrees with the late-time imbalance of the simulation, 0.9965. The imbalance that stays near unity in \figpanel{fig:analog_simulation}{f} is thus the equilibrium value at the energy of the initial state, not a signature of localization, whereas sampling the infinite-temperature ensemble~\cite{thomsonUnravellingQuantumDynamics2024} would predict that it vanishes. Infinite-temperature correlators therefore describe quenches from typical initial states, such as the resonant states here, but not experiments that prepare low-energy or negative-temperature states.

\end{document}